2018

# Real-time SIL Emulation Architecture for Cooperative Automated Vehicles

Nitish Gupta
*University of Central Florida*

REAL-TIME SIL EMULATION ARCHITECTURE FOR COOPERATIVE AUTOMATED VEHICLES

by

NITISH A. GUPTA
B.E. University of Mumbai, 2014
M.S. University of Central Florida, 2018

A thesis submitted in partial fulfillment of the requirements
for the degree of Master of Science
in the Department of Computer Engineering
in the College of Engineering & Computer Science
at the University of Central Florida
Orlando, Florida

Summer Term
2018

Major Professor: Yaser P. Fallah

# ABSTRACT

This thesis presents a robust, flexible and real-time architecture for Software-in-the-Loop (SIL) testing of connected vehicle safety applications. Emerging connected and automated vehicles (CAV) use sensing, communication and computing technologies in the design of a host of new safety applications. Testing and verification of these applications is a major concern for the automotive industry. The CAV safety applications work by sharing their state and movement information over wireless communication links. Vehicular communication has fueled the development of various Cooperative Vehicle Safety (CVS) applications.

Development of safety applications for CAV requires testing in many different scenarios. However, the recreation of test scenarios for evaluating safety applications is a very challenging task. This is mainly due to the randomness in communication, difficulty in recreating vehicle movements precisely, and safety concerns for certain scenarios. We propose to develop a standalone Remote Vehicle Emulator (RVE) that can reproduce V2V messages of remote vehicles from simulations or from previous tests, while also emulating the over the air behavior of multiple communicating nodes. This is expected to significantly accelerate the development cycle. RVE is a unique and easily configurable emulation cum simulation setup to allow Software in the Loop (SIL) testing of connected vehicle applications in a realistic and safe manner. It will help in tailoring numerous test scenarios, expediting algorithm development and validation as well as increase the probability of finding failure modes. This, in turn, will help improve the quality of safety applications while saving testing time and reducing cost.

The RVE architecture consists of two modules, the Mobility Generator, and the Communication emulator. Both of these modules consist of a sequence of events that are handled based on the type of testing to be carried out. The communication emulator simulates the behavior of MAC layer while also considering the channel model to increase the probability of successful transmission. It then produces over the air messages that resemble the output of multiple nodes transmitting, including corrupted messages due to collisions. The algorithm that goes inside the emulator has been optimized so as to minimize the communication latency and make this a realistic and real-time safety testing tool. Finally, we provide a multi-metric experimental evaluation wherein we verified the simulation results with an identically configured ns3 simulator. With the aim to improve the quality of testing of CVS applications, this unique architecture would serve as a fundamental design for the future of CVS application testing.

Dedicated to my parents and sister

# ACKNOWLEDGMENTS

I would like to acknowledge and thank the following important people who have supported me, not only during the entire course of this project, but throughout my Master's degree.

Firstly, I would like to express my gratitude to my supervisor Prof. Dr. Yaser Fallah, for his unwavering support, guidance and insight throughout this research project. I would also like to thank Somak Datta Gupta from Ford Motor Company. Without his clear vision and ideas, this research would not have been possible. Somak's encouragement and belief in what he does has inspired me.

And finally, I would like to thank Abhianshu, Namita and all my close friends and family. You have all encouraged and believed in me. You have all helped me to focus on what has been a hugely rewarding and enriching process.

# TABLE OF CONTENTS

# LIST OF FIGURES

# LIST OF TABLES

# LIST OF ACRONYMS

ABS        Anti-lock Braking Systems

AC         Access Categories

ACK        Acknowledgement

ADAS      Adaptive Driving Assistant Systems

AIFS       Arbitration Inter-Frame Space

API        Application Programming Interface

ASTM     American Society for Testing Materials

BC         Backoff Counter

BSM       Basic Safety Messages

CAV       Connected and Automated Vehicles

CBP       Channel Busy Percentage

CBR       Channel Busy Ratio

CCH       Control Channel

CSMA/CA   Carrier-Sense Multiple Access with Collision Avoidance

CTS       Clear-to-send

CVS       Cooperative Vehicle Safety

CW        Contention Window

DCF       Distributed Coordination Function

DIFS       DCF Inter-Frame Space

DNPW    Do Not Pass Warning

DSRC     Dedicated Short Range Communication

| | |
|---|---|
| D2D | Device-to-Device |
| ECEF | Earth-Centered, Earth Fixed |
| EDCA | Enhanced Distributed Channel Access |
| ENU | East-North-Up |
| FEC | Forward Error Correction |
| GPS | Global Positioning System |
| GUI | Graphical User Interface |
| IFS | Inter-Frame Space |
| IP | Internet Protocol |
| IRM | International Reference Meridian |
| IRP | International Reference Pole |
| FCC | Federal Communications Commission |
| FCW | Forward Collision Warning |
| FIFO | First-In-First-Out |
| LDW | Lane Departure Warning |
| LLC | Logical Link Control |
| LoS | Line-of-Sight |
| LTE | Long-Term Evolution |
| LTS | Long Training Symbols |
| MAC | Medium Access Control |
| MANET | Mobile Ad-hoc Network |
| MLME | MAC Layer Management Entity |
| NS3 | Network Simulator v3 |
| OBU | On-board Unit |
| OFDM | Orthogonal Frequency Division Multiplexing |
| PD | Propagation Delay |
| PDU | Packet Data Units |
| PER | Packet Error Rate |
| PLCP | Physical Layer Convergence Protocol |
| PMD | Physical Medium Access |
| QoS | Quality of Service |
| RTS | Request-to-send |
| RSSI | Received Signal Strength Indicator |
| RSU | Road-side Unit |
| RVE | Remote Vehicle Emulator |
| SCH | Service Channel |
| SIFS | Short Inter-Frame Space |
| SINR | Signal Interference to Noise Ratio |
| STS | Short Training Symbols |
| SUMO | Simulation of Urban Mobility |
| TXOP | Transmit Opportunity |
| UE | User equipment |
| Ups | User Priorities |
| V2I | Vehicle-to-Infrastructure |
| V2X | Vehicle-to-Everything |
| V2V | Vehicle-to-Vehicle |
| VANET | Vehicular Ad-Hoc Network |
| WAVE | Wireless Access for Vehicular Environment |
| WBSS | Wave Basic Service Set |
| WLAN | Wireless LAN |
| WME | WAVE Management Entity |
| WSE | WAVE Security Entity |
| WSMP | WAVE Short Message Protocol |

# CHAPTER 1: INTRODUCTION

In the recent years, the automotive industry has seen an evolution of Automated vehicle systems. Safety has been the main motivation behind the research and development in Automated vehicle systems. However, the recent fatal accidents involving autonomous vehicles have increased the safety concerns for the entire automotive industry. Often the causes of these accidents are the failure of one or more automated driving subsystem. Thus, robust testing and validation is an inevitable and essential aspect of developing such subsystems in automated vehicles.

Currently, automated driving systems can perceive their environment broadly in two ways, through a stack of onboard sensors or through cooperation and communication. The sensor-based perception extensively relies on local sensors such as, Lidars, radars, sonars and cameras to plan, navigate, and make safety decisions. However, vehicles with local sensors alone have a limited field of view and thus they are unaware of other vehicles' behavior at farther distances as well as in occluded spaces. On the other hand, a modern Connected and Automated Vehicle (CAV) uses a Dedicated Short-Range Communication (DSRC) [1], [2] to communicate with other vehicles. The V2V communication is based on map-sharing between automated vehicles using Basic Safety Messages (BSM) to expand the perceivable environment and thus efficiently plan longer maneuvers. The DSRC based V2V communication has fueled development of several Advanced Driver Assistance Systems (ADAS) for Cooperative Vehicle Safety (CVS) [3], [4]. Several ADAS applications such as Forward Collision Warning (FCW), Blind Spot Detection Warning, Lane Departure Warning (LDW), Do Not Pass Warning (DNPW) and so on, are already being deployed in many high-end vehicles. However, the robustness of such applications primarily relies on the quality of testing being carried during development.

## 1.1 Related Work

The most common way to test V2V safety applications is through initially through lab test followed by field tests using actual test vehicles. Analysis and testing of V2V safety applications are difficult using actual road tests, due to the cost and repeatability as well as the challenge of considering all the traffic scenarios and various communication conditions. Even when tests identify an issue, it is difficult to test the fixes due to the challenge of recreating the exact scenario. On the other hand, Vehicular Ad-Hoc Network (VANET) based Simulators such as NS3 [5], VEINS [6] and OMNET++ [7] have proved to be an economical, repeatable and easily configurable way of evaluating safety applications. Various mathematical communication models [8], [9], [10] have been developed for evaluation and analysis of ADAS using network simulators. Analysis of these highly accurate models provided insights and thus development of channel congestion control algorithms [3], [11] and [12]. Simulation model-based testing of ADAS is an efficient but not exhaustive method for evaluation of various time-critical safety applications. A real-world environment is dynamic and consists of numerous critical and unimaginable scenarios which would be otherwise impossible to model in a simulator. Although simulators are useful for initial testing of ADASs, they are clearly not sufficient for a comprehensive testing and validation of CVS applications. An early attempt to address this issue is seen in [13], which developed a BSM Emulator to help the development of advanced vehicle safety applications.

In this paper, we propose a Remote Vehicle Emulator (RVE) framework that utilizes the advantages of both, field-based as well as simulator-based testing of safety applications. RVE allows recreation of the environment as observed by a CAV. For this purpose, the host vehicle device receives the BSM messages from a cloud of virtual vehicles, in a way to emulate actual vehicles in a field test. This framework allows repetition of previous field tests, and in effect studying the performance and fine-tuning of the CAV's hardware and software.

The emulation framework supports incorporation of simulated vehicle movements, broadening the testing possibility to that of scenarios that may be too dangerous or too costly to easily reproduce in a field test. For example, near-crash scenarios or large-scale scenarios (hundreds of vehicles on the channel) are very difficult to create or recreate. Vehicle and communication traces for such expansive and complex scenarios can be generated using a combination of limited field test and simulations. The mobility traces [14] can then be fed to the RVE and host vehicle devices to reconstruct the scenario for a host vehicle CAV system. The emulated standardized over-the-air messages representing the virtual vehicles can be received by the host vehicle and to it would appear as if the emulated CAVs are actually present around it.

## 1.2 Contribution

The following are the most important contribution of this thesis.

1. **Development of BSM Rich Testing Environment:** The proposed Remote Vehicle Emulator can simulate a BSM cloud while also adhering to the underlying communication protocols. This cloud of safety data is essential for testing every corner case of any safety application.

2. **Design of Embedded Communication Emulator:** The Communication emulator uses only a single DSRC device to simulate thousands of virtual devices behavior in real-time. Moreover, the simulated packets are sent over the air in a manner to imitate a realistic wireless vehicular network without requiring any other external component.

3. **Incorporated Mobility Handler:** The RVE design has a built-in Mobility Handler, which can generate several mobility traces without any hassle using real or simulation vehicle movements.

4. **Multi-metric Validation:** The validation of emulator design has shown a high confidence in the simulation behavior.

## 1.3 Organization of the Thesis

Understanding the design of RVE requires the fundamentals of various networking and communication standards. Chapter 2 presents all the related background details required to understand the underlying nuance of the RVE architecture. This chapter consists of an overview of different types of Vehicular networks. Later we discuss the WAVE reference model and the various protocols associated with it. The WAVE discussion leads us towards the Medium Access Control and Physical layer of IEEE 802.11p, which explains the operation of different types of MAC coordination schemes. This is followed by a brief overview of types of channel losses and their impact on the receiver model.

Chapter 3 focuses on the design of RVE, where we go through the two sub units of the emulator design, Mobility Log Generator and Communication Emulator (CommSim). Then in Chapter 4 the implementation of the emulator design is presented in detail. Various features of CommSim are discussed in this chapter such as Branch operations, the design of receiver model, logger and more. Chapter 4 also provides the configuration details related to CommSim as well as Network Simulator 3 (NS3) which would be later required for experimental evaluation.

Chapter 5 is dedicated to the experimental evaluation of RVE and provides an overview of the selected evaluation metrics followed by a thorough validation of CommSim by comparing the simulation results with NS3. Finally, this chapter presents an evaluation of the receiver model design.

The final Chapter provides the conclusion and proposes future work.

# CHAPTER 2: BACKGROUND

## 2.1 Vehicular Communication

### 2.1.1 Dedicated Short Range Communication

In early 90s, many industries were focused on developing the road toll collection devices based on RFID transponders. This gave rise to the development of a few road safety applications on 950 MHz channel. It was evident that a concrete network architecture was required to develop and deploy these vehicle safety applications. This led to the development of a new standard named Dedicated Short-Range Communication (DSRC) introduced particularly for vehicular safety applications. DSRC is an adoption of IEEE 802.11 and IEEE 1609.x standard suites. DSRC uses IEEE 802.11 in a broadcast mode to have minimum latency. Further development in DSRC enabled communication between Vehicle-to-Infrastructure in addition to Vehicle-to-Vehicle.

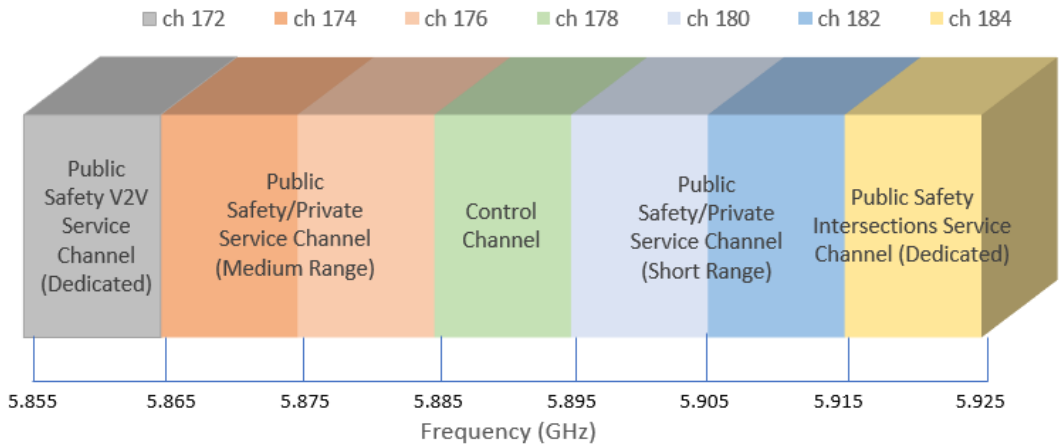

Figure 2-1: DSRC Frequency Spectrum

Source: Volker Fessmann, NHTSA, 5.9 GHz DSRC Spectrum and Potential Band Sharing

DSRC uses 2-way Line-of-Sight (LoS) communication which makes it low-cost, short-range and high bandwidth network. Traditional Mobile Ad-hoc Networks (MANETs) employ an intermediate station or satellite for communication. However, DSRC is decentralized which makes vehicles to communicate with each other without a need of intermediate station. Although this standard utilizes the existing standards, it is not meant to replace the other wireless networks but is dedicated for ensuring the vehicular safety. In US, the spectrum ranging from 5.850 to 5.925 band is assigned to DSRC by Federal Communications Commission (FCC).

The maximum transmission range (~ 350 m) of DSRC is far less than other wireless technology like WiMAX and Satellites. Being a short-range LoS communication, DSRC can reach a data rate of about 27Mbps which is lower compared to cellular, WiMAX or satellite communication. In vehicular applications, a fast communication is desired between speeding vehicle while also maintaining data privacy. Thus, DSRC has a strict Quality of Service (QoS) requirement which can ensure safety and anonymity.

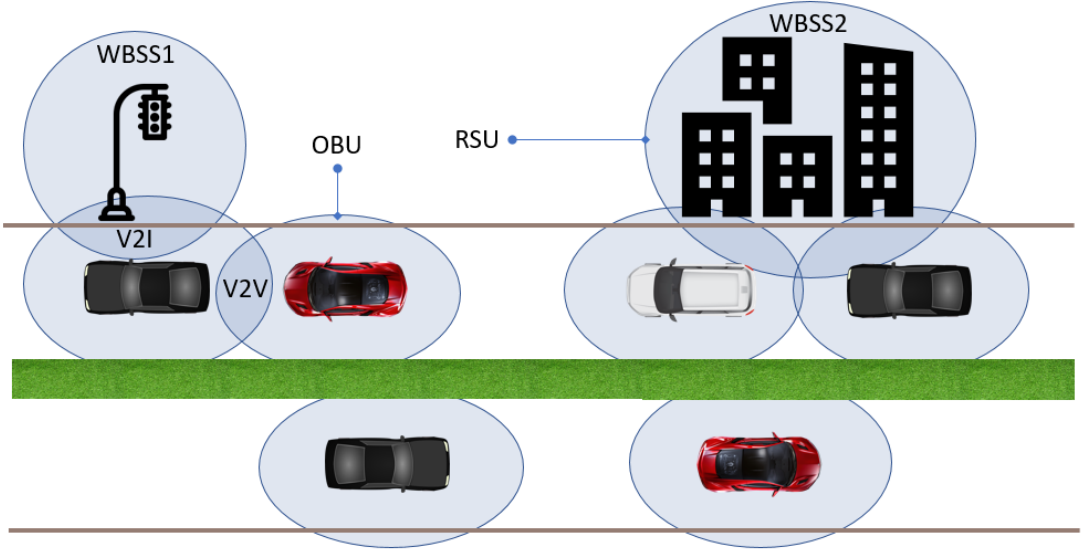

Figure 2-2: RSU, OBU and WBSS

### 2.1.1.1 DSRC Devices

There are two basic units of DSRC – Road-side Unit (RSU) and On-board Unit (OBU). RSUs are connected to infrastructure and is stationary while OBUs are connected to the moving vehicles. OBUs connects to both external as well as internal vehicle network to share information obtained through local sensors. Each RSU consists of a unique Wave Basic Service Set (WBSS) identifier, which divides the road into RSU communication zones. OBU equipped moving vehicles can connect to one RSU at a time using the WBSS while also maintaining the communication with other vehicles. Thus, each OBU gets limited time to connect to a WBSS, exchange safety information and then leave the zone. The exchange of safety information is performed using Wave Short Message Protocol (WSMP). The Wave standard offers two separate channels based on the type of information exchange namely, Control Channel (CCH) and Service Channel (SCH). The CCH is used for the exchange of Safety critical messages, while the SCH is used to exchange information based on commercial applications like Nearest Gas station, Credit Card transaction, Resting zones and so on.

### 2.1.1.2 Basic Safety Messages

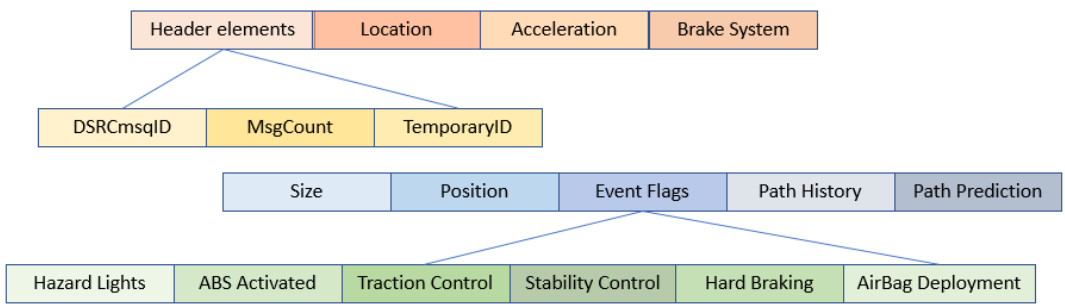

Figure 2-3: BSM Field Format

Source: Brian Cronin, ITS, Vehicle Based Data and Availability

In DSRC, the safety information shared between Vehicle and/or Infrastructure for enabling the cooperative safety in encoded into a special type of packet called Basic Safety Messages. The BSM message standard format is defined under SAE J2735 standard. The BSM is customized for local broadcasting and low latency and consists of two parts. The BSM part 1 contains the core vehicle safety information such as position, heading, speed, acceleration, vehicle dimensions and brake status. The BSM part 1 information is broadcasted approximately 10 times per second. This ensures that a vehicle receives the most updated information about the status of its neighboring vehicles which is essential for critical safety applications. On the other hand, BSM part 2 consists of less safety critical information such as related to wipers, exterior lights status, Anti-lock Braking Systems (ABS), ambient pressure and/or temperature and rain sensors. This type of information is shared within BSM in case of special events such as ABS activation. In addition, this information is customizable depending upon the type of situation and vehicle models. Since BSM part 2 information is not safety critical, it is shared less often as compared to BSM part 1.

### 2.1.2  *Cellular Long-Term Evolution*

The 3GPP release 14 introduced a support for enhanced vehicular safety applications. C-V2X can operate in a non-Line-of-Sight field with low latency and high throughput, providing comparable performance to DSRC standard and aiming to replace it. The previous releases based on Device-to-Device (D2D) communication were network dependent where User equipment (UE) could communicate with each other using an intermediate evolved NodeB (eNodeB). However, rel. 14 operates using PC5 interface and is designed to be network independent. This technology is still under development at the time of this writing.

## 2.2 WAVE Reference Model

The Wireless Access for Vehicular Environment (WAVE) model design is based on the existing IEEE 802.11 and IEEE 1609.x standards along with some newly introduced features for vehicular safety applications [2]. The model is specifically enhanced for handling communication between high speed moving vehicles (V2V) and infrastructure (V2I). As moving vehicles are required to connect or disconnect to a remote network with minimum latency, the reference architecture only operates in the broadcast mode of IEEE 802.11 standard.

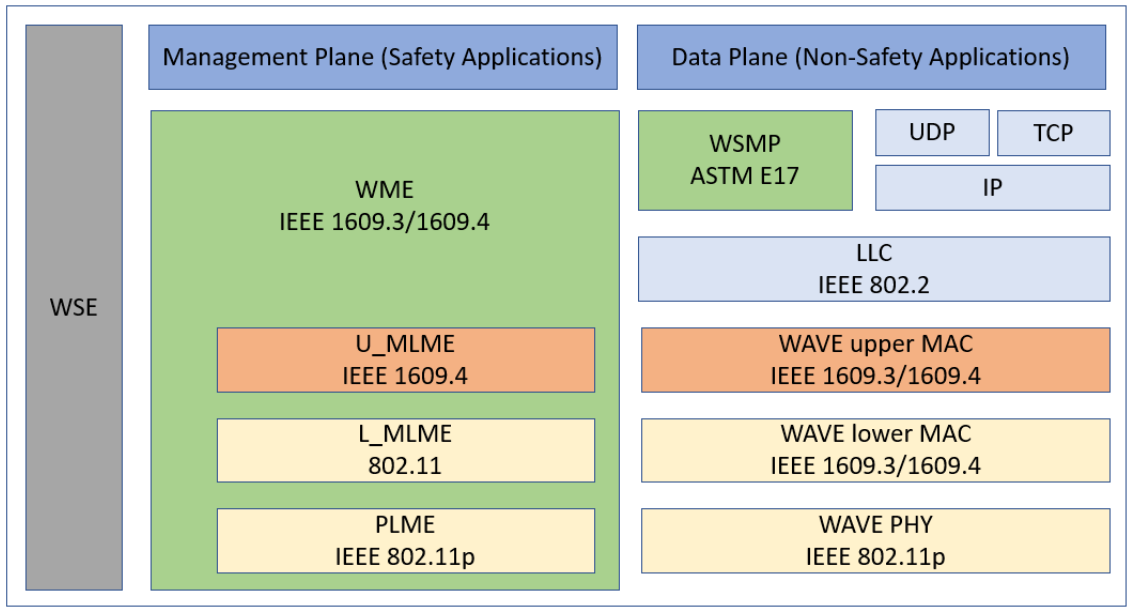

Figure 2-4: WAVE Reference Model [2]

Let us now discuss each component of the Wave reference model in more detail. As shown in Figure 2-4, the Wave architecture can be roughly divided into two sub-planes namely Management Plane and Data Plane [18]. The main responsibility of the Management plane includes synchronization, switching channels and so on. On the other hand, Data plane is responsible for handling data such as adding/removing frame headers and passing the data to the upper or lower layers.

The WAVE Management Entity (WME) based on IEEE 1609.3/4 is a WAVE architecture specific component. Whenever a packet is scheduled for transmission, various transmission parameters and QoS priorities must be defined. Latency of a transmitted packet is of importance as the packet may contain a high priority safety message or a lower priority service message. Along with priorities it is important to specify the packet generation rate (usually 100 milliseconds for safety messages) or transmission power which decides how far the packet would travel in the channel. Thus, for different applications, WME is responsible for registering parameters like priorities, data generation rate, and transmission power.

Another unique WAVE specific networking service component is the WAVE Short Message Protocol (WSMP) that is based on American Society for Testing Materials (ASTM) E17.51 standard [1]. It is evident that the Safety messages must be transmitted with higher priority and low latency. WSMP protocol is designed specifically for generating packets for safety applications. These packets are optimized for a lower channel interference and lower latency as compared to Internet Protocol (IP) generated packets. Based on the type of message, WSMP sets the packet parameters such as channel, transmission power, data rate and so on. In addition to WSMP, WAVE networking services also supports IPv6 based datagrams.

The WAVE Security Entity (WSE) component comes from the IEEE 1609.2 standard is responsible for security management of the transmitted packets. It is essential to encrypt the safety and service messages to avoid an attack on the network. Along with Data Encryption and Key management, WSE also enforces various security policies and actively responds in case of a possible attack.

To simulate the behavior of IEEE 802.11p protocols, it is necessary to understand the behavior of MAC and PHY layers. The MAC layer is mainly responsible for managing the channel access and packet transmission over a shared network in a fairly distributed manner. Also, for safety critical applications, latency must be minimum. Thus, the MAC protocol is required to utilize the available bandwidth efficiently while providing the desired QoS to ensure reliability of control and service information. One of the basic medium access mechanisms specified by IEEE 802.11 standard based on channel contention is the Distributed Coordination Function (DCF). However, DCF is only designed to provide a best effort packet transmission and doesn't consider the QoS. In order to get a QoS, IEEE 802.11e specified a new medium access mechanism known as Enhanced Distributed Channel Access (EDCA). Let us now discuss each of these medium access categories in detail.

*2.2.1 Medium Access Control*

MAC layer specifies protocols for coordination between nodes wanting to transmit packets over a shared channel at the same time. An efficient coordination mechanism minimizes the packet collision over the channel and improves the channel utilization at the same time. When the channel is not fully observable, a problem occurs two nodes hidden from each other start transmitting to a common visible node which results in collision. This problem is commonly known as Hidden node problem. In this section we will discuss the two main channel access protocols specified under IEEE 802.11 and IEEE 802.11e namely, DCF and EDCA. Later we will discuss the Hidden node problem and its possible solutions.

2.2.1.1  Distributed Coordination Function

Distributed Coordination Function is the primary and mandatory MAC protocol specified in IEEE 802.11 standard [21]. It is essential to understand Carrier Sense Multiple Access with Collision Avoidance (CSMA/CA) based DCF before getting into EDCA.

The MAC layer maintains a transmission First-In-First-Out (FIFO) queue. For vehicular applications, only the most recent and updated information like position, speed, heading etc. is relevant. Thus, in the MAC layer of IEEE 802.11p, a transmission queue to hold a single packet is sufficient. The single element queue is overwritten each time a new packet arrives from the upper layer.

In CSMA/CA, each node with a packet to transmit first senses the channel status as busy or idle. If the node senses the channel busy, the MAC waits until the channel becomes idle again followed by an additional time interval known as the DCF Inter-Frame Space (DIFS). However, if the channel is sensed as idle after DIFS is elapsed, the MAC initiates a back-off process. To initiate the back-off, the respective MAC randomly selects a back-off counter value from a predefined Contention Window. If the channel stays idle after the DIFS deference, the node begins decrementing the Backoff Counter (BC) by one for each idle time slot. During the back-off process, if another node gets channel access before the first node could complete the count, the first node freezes the BC and waits till the channel is idle again. Now the next time it finds the channel idle, instead of picking up a new backoff value, it waits for another DIFS and continues the BC decrement from where it stopped the last time. This ensures those deferred nodes with a back-off history have higher priority over the new back-off nodes as they are only required to wait for the time equivalent to the remaining BC value.

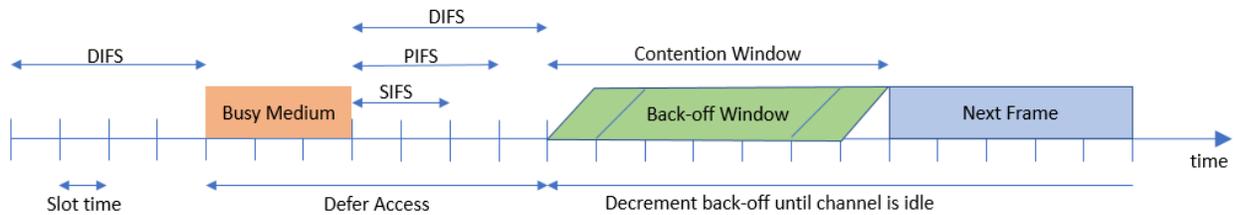

Figure 2-5: Distributed Coordination Function Timings [15]

The MAC layer of each node has a Contention Window (CW) from which it randomly selects a back-off counter value. The BC integer is picked from a Uniform distribution between the interval [0, CW]. Larger the size of CW, higher is the resolution of back-off mechanism. Higher resolution ensures that the probability of picking the same number by multiple nodes is low. However, it also implies that the wait time for certain nodes will increase. Thus, smaller CW size has shorter delays. Figure 2-5 illustrates the channel access timings of DCF. In IEEE 802.11 standard, whenever a transmitted packet is successfully received at the receiver, an acknowledgement (ACK) frame is sent back into the channel. However, this ACK frame is sent after a Short IFS (SIFS). The SIFS is always shorter than DIFS which ensures the transmission of the ACK without interfering with the CW of other nodes. Initially the size of contention window is set to CWmin, if a transmitted packet is lost and an ACK for the same isn't received, the size of the CW is doubled. The size is increased upon each successive unsuccessful transmission up to the upper limit of CW i.e. CWmax. In case when the packet gets through the channel successfully in one of the attempt, the size of CW is again reset to the initial CWmin.

2.2.1.2 <u>Enhanced Distributed Coordination Access</u>

The Enhanced Distribution Coordination Access (EDCA) protocol was designed to support QoS based access control applications. EDCA provides a distributed and distinct QoS based contention mechanism. In EDCA, the high-priority data gets a higher probability of getting transmitted as compared to the lower priority data. This is accomplished by replacing the fixed DIFS with Arbitration Inter-Frame Space (AIFS) which uses different values for each type of traffic. For instance, the traffic with higher priority get a smaller value of AIFS and it increases as we move towards lower-priority traffic data. The exact values of AIFS for different Access Categories (AC) depends on the physical layer of the model. As shown in Figure 2-6, 802.11 provides 8 different User Priorities (Ups) based on 4 ACs.

|  | | 802.1p | | 802.11e | |
| --- | --- | --- | --- | --- | --- |
| Priority | Priority Code Point (PCP) | Acronym | Traffic Type | Access Category (AC) | Designation |
| Lowest | 1 | BK | Background | AC_BK | Background |
|  | 2 | -- | Spare | AC_BK | Background |
|  | 0 | BE | Best Effort | AC_BE | Best Effort |
|  | 3 | EE | Excellent Effort | AC_BE | Best Effort |
|  | 4 | CL | Controlled Load | AC_VI | Video |
|  | 5 | VI | Video | AC_VI | Video |
|  | 6 | VO | Voice | AC_VO | Voice |
| Highest | 7 | NC | Network Control | AC_VO | Voice |

Figure 2-6: User Priorities for different Access Categories
Source: IEEE 802.11e-2005, Wikipedia

In EDCA, the bounded time for which a node is granted a contention-free access is known as Transmit Opportunity (TXOP). During TXOP, a node can send as many packets as it can for the duration of granted TXOP as long as the packet length doesn't exceed the maximum duration of the TXOP. In case the packet length is longer than the available TXOP, the frame is fragmented into smaller frames. This avoids the problem where a lower priority station acquires the access for indefinite amount of time in case of DCF. In EDCF, the fixed DIFS is replaced with AIFS which depends up on the AC and is given as,

$$AIFS[AC] = SIFS + AIFSN[AC] \times slot\_time \qquad (2\text{-}1)$$

$$where, AIFSN[AC]: represents\ Arbitration\ Inter\ Frame\ Space$$

$$slot\_time: smallest\ fraction\ of\ time$$

Figure 2-7 shows the timing relationship in EDCA. The AP announces the values of AIFSN and CW of each AC through beacon frames. In case when multiple AC senses the channel as idle and their BC is zero, there is a possibility of Collison between packets from different ACs. However, this scenario is prevented by inclusion of internal and external contention handler. The contention handler only allows the AC with higher priority to transmit over the channel while the other AC with lower priorities are required to backoff with an increase CW value.

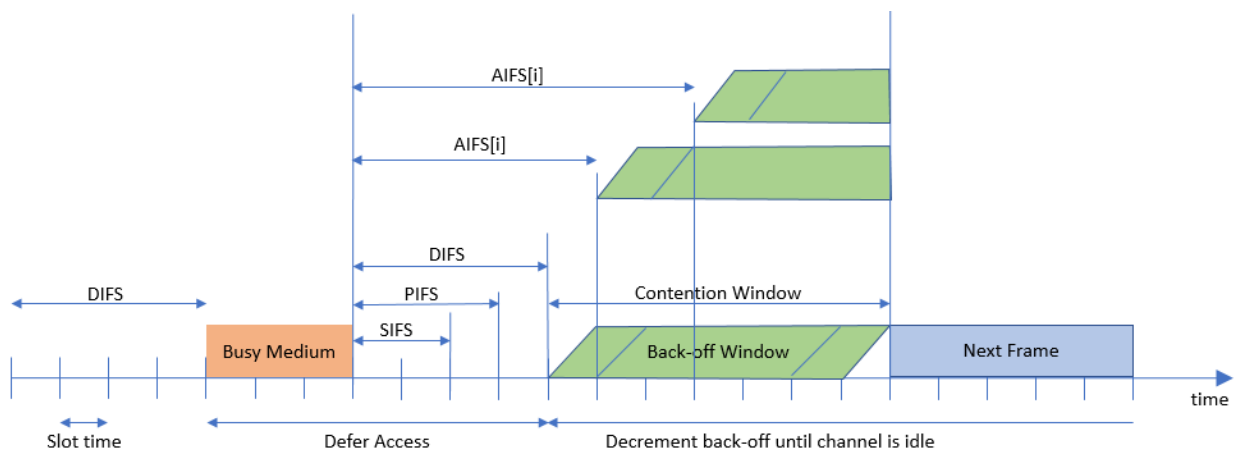

Figure 2-7: Enhanced Distributed Coordination Access Timings [15]

As discussed in section 2.2, IEEE 802.11p specifies two types of traffic, IPv6 and WSMP each with different QoS requirement. Since WSMP packets are used to communicate vehicular safety information, they get a higher priority over IP based packets which are mainly consist of service based applications. The WSMP packets contain information like channel number, data rate, transmission power and internal message priority. On the other hand, IPv6 packets similar information except that they are required to inform about their profile to MAC Layer Management Entity (MLME). The IEEE 802.11p standard supports 4 queues for each CCH and SCH access channel. Figure 2-8 shows the Channel Coordination Function specified under WAVE MAC layer. The frames from Logical Link Control Layer (LLC) are directly sent to the channel router which segregates the frames based on the type of access channel (CCH/SCH). These frames are then internally routed to one of the four access category queues depending on their priorities.

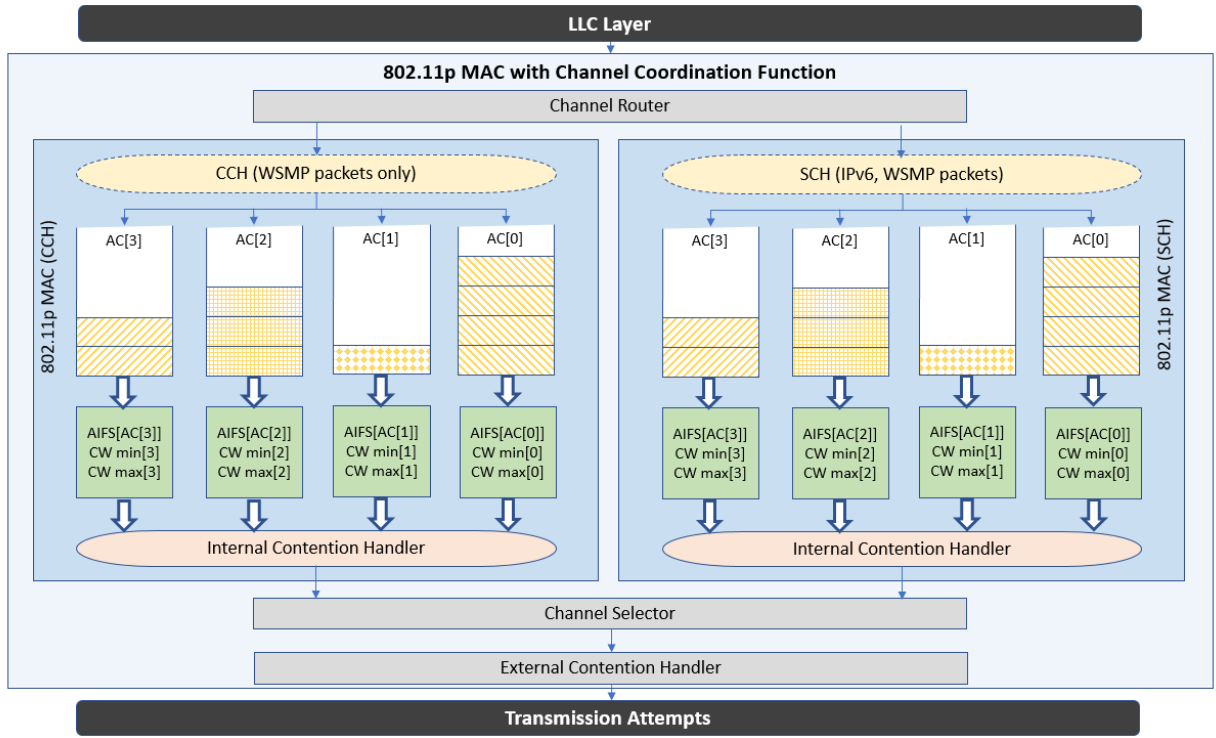

Figure 2-8: Enhanced Distributed Coordination QoS Priorities Queues [2]

## 2.2.1.3 Carrier Sense Multiple Access with Collision Avoidance

Carrier Sense Multiple Access with Collision Avoidance is a medium access control (MAC) algorithm specified by the wireless IEEE 802.11 standard [WIFI]. CSMA/CA relies on the best effort scheme to avoid collision in wireless networks. In CSMA, the node first senses the medium and transmits the messages only if the medium is idle. CSMA/CA is like CSMA except that the small control messages are exchanged before sharing the actual data, thus helps in avoiding the collision. CSMA/CA works by sharing the control frames such as Request-to-send (RTS) and Clear-to-send (CTS) between the nodes. RTS frame includes the receiver address and the transmitter address along-with the duration the transmitter needs to be in constant connection with the receiver. CTS frame includes the duration and the receiver address.

CSMA/CA works as follows: The sender sends the RTS frame for the targeted receiver. After receiving the RTS frame, the receiver sends the CTS frame to all the nodes which are in the communication range of the receiver node. All the nodes except the transmitter node stops transmitting any frame for the duration mentioned in CTS frame. Also, the transmitter node sends the actual data message to the receiver node after receiving the CTS frame. The receiver node sends the ACK including the duration and receiver address guaranteeing that the message is received successfully at the receiver's end.

CSMA/CA is efficient protocol when the data is large as it lowers the collision cost with the use of small control messages. Also, it helps in avoiding the hidden node problem by virtually sensing the medium with the help of RTS/CTS frames.

In IEEE 802.11, CSMA/CA operates in two modes – base station and ad-hoc network. However, as vehicular communication is time-critical, the presence of these additional frames would increase communication latency. Thus, IEEE 802.11p has only adopted the broadcast mode where nodes communicate with each other without the control frames.

2.2.1.3.1 Propagation Delay Problem

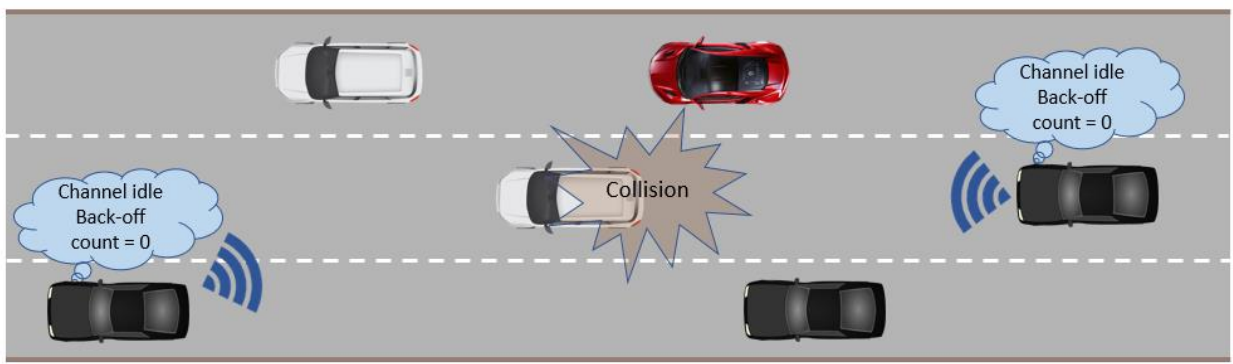

Figure 2-9: Propagation Delay Problem

The transmission of a signal through the channel takes variable amount of time based on the distance between the transmitting and receiving nodes. This delay contributes to a significant amount of packet losses in the channel. In a fully connected CSMA/CA network, a node starts transmitting whenever it senses the channel as idle and its back-off counter is zero. However, in a dense network, it is possible that another node at a certain distance also finds the channel idle with its BC zero at the same time. Thus, both nodes start their respective transmission which would eventually result into a collision and a damaged packet will be received by other nodes. The collisions due to signal propagation delay contributes to channel utilization and the Packet Error Rate (PER). Thus, it is important to consider the effect of propagation delay in simulation of CSMA/CA. To address this situation, an additional fixed time interval called PD is when a node initiates its transmission. During PD, all those such nodes that contribute to this effect are identified and an extended damaged packet is sent over the channel indicating that a collision has occurred due to signal propagation delay. The length of the damaged packet is equal the union of first and last overlapped packet.

2.2.1.3.2  Hidden Node Problem

Hidden Node Problem is a very common problem in wireless networks such as MANETs and VANETs. Figure 2-8 illustrates the hidden node problem in wireless communication. This problem arises when two Access Points (A1 and A2) are not within the communication range of one another but both Access Points can see a third Access Point (A3). So, A1 can communicate with A3 and A2 can communicate with A3. If A1 and A2 try to communicate with A3 at different times, then the transmitted messages are received without any interruption. However, if both the Access Points A1 and A2 desires to communicate with A3 at the same time, then the messages received by Access Point A3 is not interpretable because the two messages intermingle near the peripheral of A3. This is due to the fact that Access Points A1 and A2 cannot sense that the medium is busy as they are out of each other's communication range. Thus, A1 is hidden to A2 and A2 is hidden to A1 while communicating with A3. Therefore, A1 and A2 cannot conclude that their messages are colliding with each other and continue transmitting their messages oblivious to the fact that A3 is unable to interpret their garbled messages. This situation is known as Hidden Node Problem.

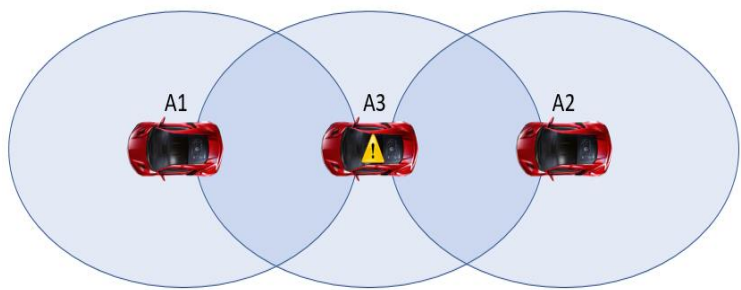

Figure 2-10: Hidden Node Problem

There are various solutions proposed to solve Hidden Node Problem. One main solution to resolve Hidden Node Problem is the introduction of Request-to-send/Clear-to-send (RTS/CTS) acknowledgement. This type of handshake starts when Access Point A1 sends the RTS to Access Point A3 and A3 acknowledges to all the Access Points which are in A3's communication range that A3 is going to communicate with A1. So, A3 will wait till the communication is going on. However, the inclusion of RTS/CTS incurs an additional over-head on the transmission by increasing the transmission time. So, this solution partially solves the hidden node problem. Another solution to overcome this problem is to increase the transmission range of the access points so that most of the nodes can see each other. However, if the transmission range increases, then the number of nodes for communication also increases which further increases the hidden node problem for the additional new nodes. Thus, a single solution is unable to solve the hidden node problem completely. Some other solutions to overcome this problem are omni-directional antennae, centralized access given to a master node like in protocols such as token passing.

### *2.2.2 Physical Layer*

The PHY layer of IEEE 802.11p is adapted from the IEEE 802.11a standard. The PHY layer operates in a narrower bandwidth of 10 MHz which is exactly the half of the available channel bandwidth in 802.11a. In effect, the subcarrier spacing is halved while the symbol length is doubled from 4 µs in 802.11a to 8 µs in 802.11p. This basically doubles the entire timings of Orthogonal Frequency Division Multiplexing (OFDM) parameters involved as compared to the 802.11a. This standard also specifies an arrangement of total 64 subcarriers of which 48 subcarriers are used for data, 4 as pilots and 12 as guard band.

The PHY layer is also an interface between the MAC layer and the medium for the transmission or reception of the packets. Based on IEEE 802.11a, 802.11p PHY also has two main sub-layers namely Physical Layer Convergence Protocol (PLCP) and Physical Medium Access (PMD). The PLCP is responsible for communication between PHY and MAC layer. MAC layer and above layers generate Packet Data Units (PDU) which are then transformed and processed to converge into an OFDM frame. These OFDM frames are then transmitted through the medium such as radio channels or transmission lines.

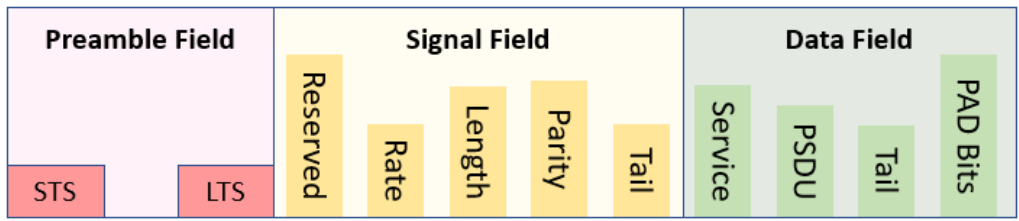

Figure 2-11: IEEE 802.11p PHY PDU Frame Format [2]

The PDUs received by the PHY layer from the upper layer consists of three types of fields namely, Preamble, Signal and Data field. As shown in Figure 2-11, the preamble field consists of Short Training Symbols (STS) and Long Training Symbols (LTS) for specifying the frequency channel behavior. The signal field which is based on the type of message to be transmitted is obtained from the WME register and added by the WSMP layer of the WAVE model. This field consists of information like packet rate, message length, parity and tail. This field also provides the information related to modulation scheme and error coding. Finally, the application data is store in the following data field.

## 2.3 Pathloss, Shadowing and Multipath Fading

Wireless Networks or WLANs are highly susceptible to interference, noise, and other channel impediments. Thus, the transmitted signals suffer a loss of power as a function of distance while traveling through the lossy medium. These impediments are unpredictable and change with the node movement and the dynamics of environment. In wireless communication, the channel impediments can be broadly classified into three types of effects namely, propagation path loss, shadowing and multi-path fading. The path loss and shadowing contribute to a large-scale signal loss and is generally deterministic in nature. On the other hand, multi-path fading has a small-scale effect on the signal propagation and can only be represented statistically. These losses are usually represented in decibel (dB) unit.

The path loss commonly occurs due to the attenuation of signal as it propagates through the channel. in the most LoS applications called free space propagation loss. Free space path loss can be represented using a simple deterministic model which is a function of intrinsic properties of the antenna used, atmospheric absorption and distance of the receiver. Free space path loss model has its main application in Satellites and LoS radio links communication.

In vehicular network, an on-road vehicle travels through spaces occupied by buildings, intersection, pedestrians, other vehicles and so on. These surrounding objects contributes to various types of signal impediments like reflection, diffraction and scattering. The combined effect of these three phenomena on the signal propagation is known as Multipath fading. The loss due to reflection occurs when the signal falls on an object larger than its wavelength. Diffraction is a result of signal obstruction due to objects with sharp surfaces or irregularities. Scattering occurs when the medium is densely occupied with objects that have smaller reflective surface as compared to the signal wavelength. Multipath fading is non-deterministic and complex to approximate due to the dynamics of environment. Several models have been proposed to mitigate the effect of multipath fading in vehicular communication such as Multi-ray models, Okumura model, Hata model, COST 231 etc. Figure 2-12 shows different sources of Multipath fading.

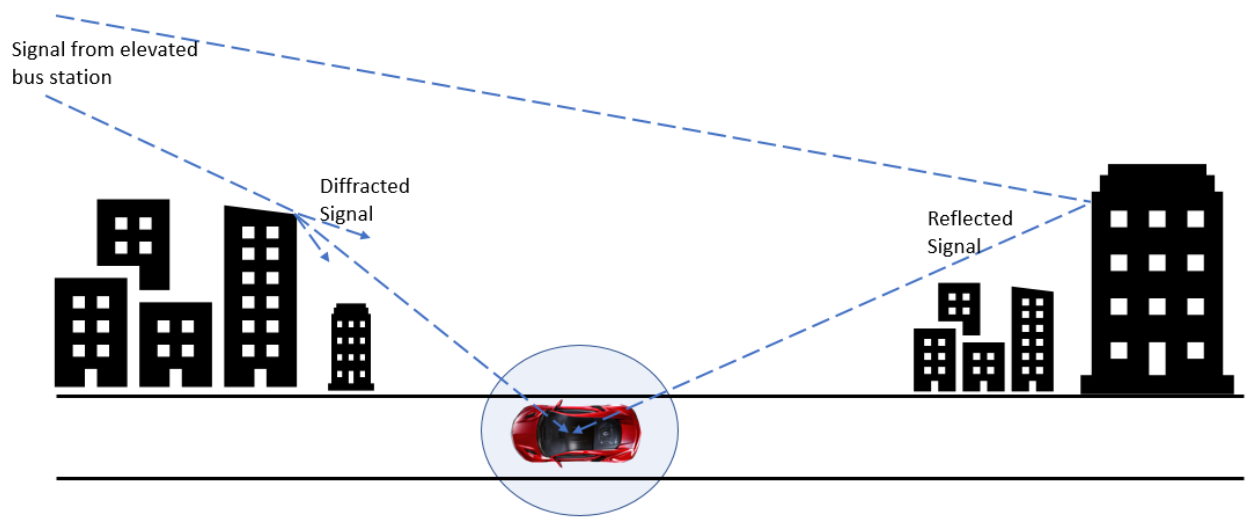

Figure 2-12: Multipath Fading

Signal propagation is also affected when there are obstacles between the transmitter and receiver. This effect is known as Shadowing. This results in random variation on local mean powers due to the interference of obstacles in signal path. Deterministic models can be used to approximate the physical underlying model.

## 2.4 Vehicular Safety Applications

With the advancements in vehicular network, several vehicular safety applications have emerged to increase the active road safety. Each safety application consists of handling a unique scenario where depending up on the communication between vehicles a series of events occur and several decisions are made. In addition to vehicular communication some applications also rely on local vehicle sensor like cameras to understand and make inferences based on the scenario and sensor data. The driver of the vehicle is alerted in case of a possible hazardous situation. Let us briefly discuss a few critical safety applications in the next subsections.

### *2.4.1  Forward Collision Warning*

The Forward Collision Warning (FCW) application is designed to warn the driver of the Host Vehicle (HV) in a situation where a rear-end collision might occur with a Remote Vehicle (RV) travelling ahead in the same lane, if the braking isn't applied immediately. This is one of the most common scenario which mostly occurs due to distracted driving.

### *2.4.2  Blind Spot Warning / Lane Change Warning*

The blind-spot zone of a vehicle is the part of road that isn't observable from inside the vehicle using any mirror. The purpose of Blind-Spot Warning (BSW) or Lane Change Warning (LCW) application is to warn a driver about the presence of a RV in the blind-spot zone of the HV. Thus, while changing lanes, the driver will be alerted if there is another vehicle in the blind-spot travelling in the same direction and but on the adjacent lanes.

# CHAPTER 3: EMULATOR DESIGN

On a high level, RVE consists of two main units, Mobility Log Generator and Communication Emulator. Mobility Log Generator is responsible for generating emulated vehicle trajectories (refer to Figure 3-1). These trajectories are logged and then fed to the communication emulator module. Communication Emulator simulates MAC layer behavior among the emulated vehicles and creates a cloud of virtual vehicles around actual vehicles. Each emulated vehicle in the cloud communicates with other emulated and actual vehicles based on the underlying communication protocols and follows the same trajectory as provided by the log generator unit. Let us now discuss each of the two units mentioned above in more detail.

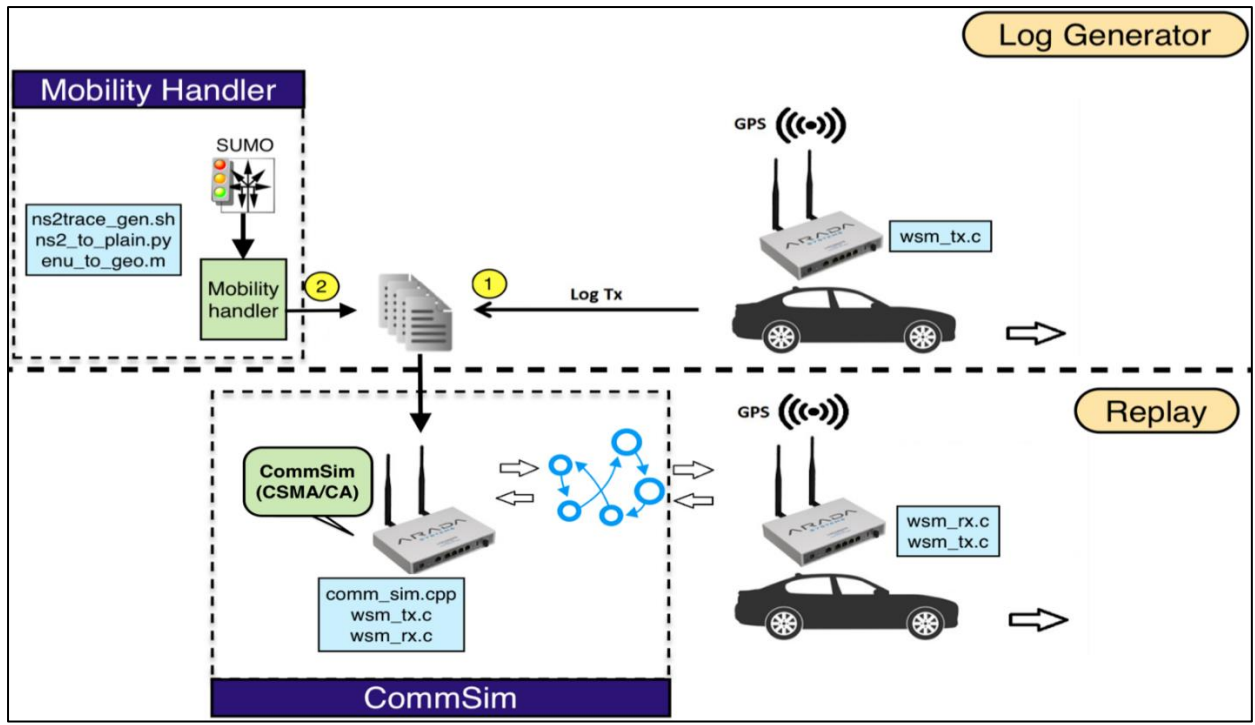

Figure 3-1: Remote Vehicle Emulator Architecture

3.1     Mobility Log Generator

The emulated vehicles are required to follow a predefined trajectory while the safety applications are being tested. We thus need a method to generate these trajectory logs which can be then fed to the communication emulator. As shown in Figure 3-2, these trajectory logs can be generated in two ways - recorded using real-vehicles or simulated using a mobility handler. Logs generated from actual vehicles are highly accurate but also time-consuming due to on-field experiments. Moreover, testing every corner case of safety application require more traffic in which case generating mobility logs from real vehicle turns out to be an inefficient and costly experiment setup.

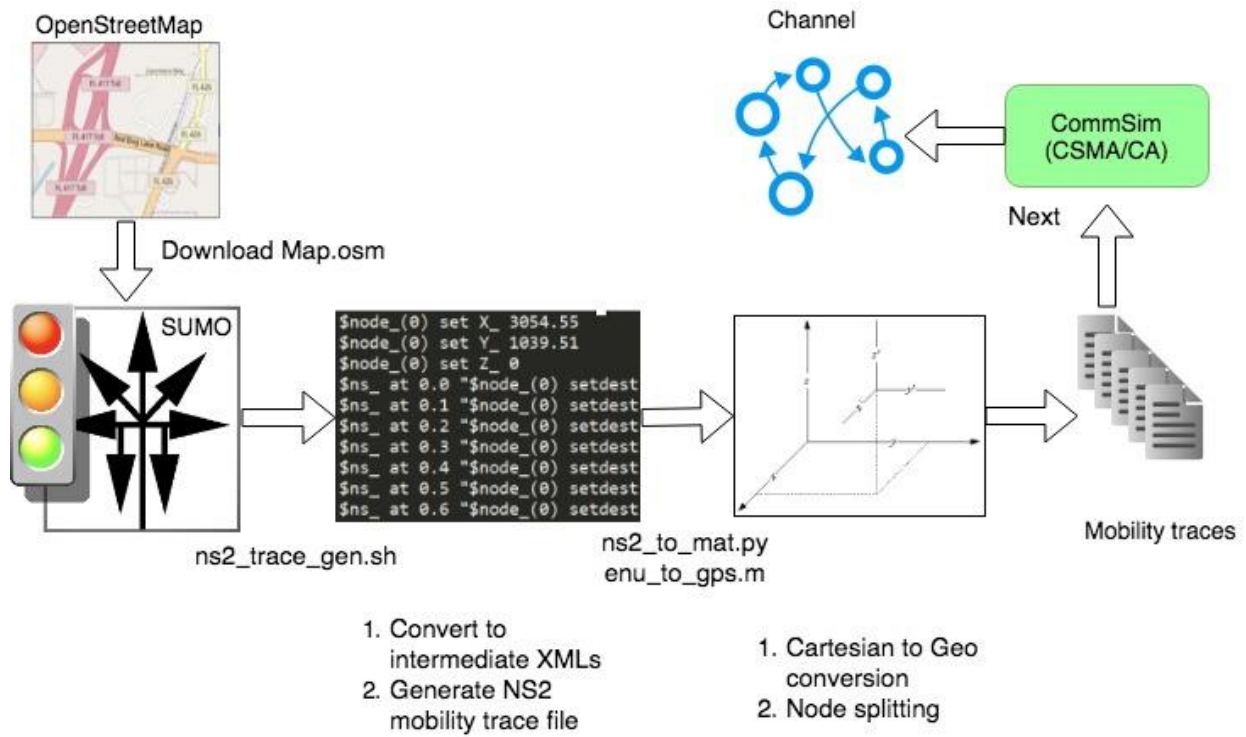

Figure 3-2: Mobility Handler Flow Chart

On the other hand, mobility handler employs a traffic simulator called SUMO [14] to generate simulated trajectory logs. The location on the map where safety applications are being tested can be exported from a map API (in our case OpenStreetMap) and fed directly to the SUMO simulator. Based on the vehicle density and road scenario, SUMO generates a ns2 mobility log. This mobility log contains the map-sharing information such as timestamp, vehicle ID, GPS position (local Cartesian coordinates), speed, heading of the vehicle and so on. The fidelity of these movements would be as good as that of the simulator. Since SUMO generates the GPS position in Cartesian coordinates, it must be first translated to the usable geodetic coordinates followed by splitting each node into separate log files. These individual nodes' logs are then used by the Communication Emulator unit to emulate communication between them.

*3.1.1 Geographic Coordinate Conversion*

The conversion from East-North-Up (ENU) to Geodetic is a two-step process. First the local ENU coordinates are required to be converted to the equivalent Earth-Centered, Earth Fixed (ECEF) coordinate system. The relationship between these three coordinate systems can be visualized through Figure 3-3. The ECEF is a Cartesian coordinate system based geographic coordinate system. The origin point of ECEF is aligned with the center of the earth, which makes it Earth Centered. On the other hand, the axis of ECEF are fixed w.r.t. to the earth's surface and aligned with the International Reference Pole and Meridian (IRP, IRM) which defines the Earth-Fixed notion. The z-axis isn't aligned to the earth's rotational axis but extends towards the true North. As these axes are aligned and fixed with respect to the earth's surface, the axes move as the earth rotates and hence it's easier to convert the ECEF to Geodetic coordinate systems.

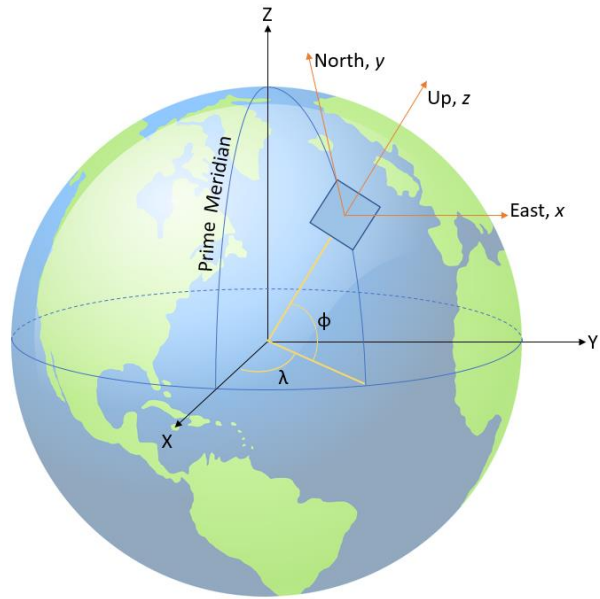

Figure 3-3: ENU, ECEF and Geodetic reference

The trigonometric conversion from ECEF to geodetic coordinate system can be solved using several methods. The Newton-Raphson and Ferrari's are the two most commonly used methods involving a series of trigonometric relations and substitutions. On the other hand, the conversion from ENU to ECEF coordinate system is comparatively simple and could be obtained by solving the following transformation,

$$\begin{bmatrix} X \\ Y \\ Z \end{bmatrix} = \begin{bmatrix} -\sin\lambda & -\sin\emptyset\cos\lambda & \cos\emptyset\cos\lambda \\ \cos\lambda & -\sin\emptyset\sin\lambda & \cos\emptyset\sin\lambda \\ 0 & \cos\emptyset & \sin\emptyset \end{bmatrix} \begin{bmatrix} x \\ y \\ z \end{bmatrix} + \begin{bmatrix} Xr \\ Yr \\ Zr \end{bmatrix} \quad (3\text{-}1)$$

*where,* $X, Y, Z$ denotes the ECEF coordinates

$x, y, z$ denotes the ENU coordinates

$\emptyset$ denotes the geodetic Latitude

$\lambda$ denotes the geodetic Longitude

An example conversion from one of the RVE execution is shown in Figure 3-4.

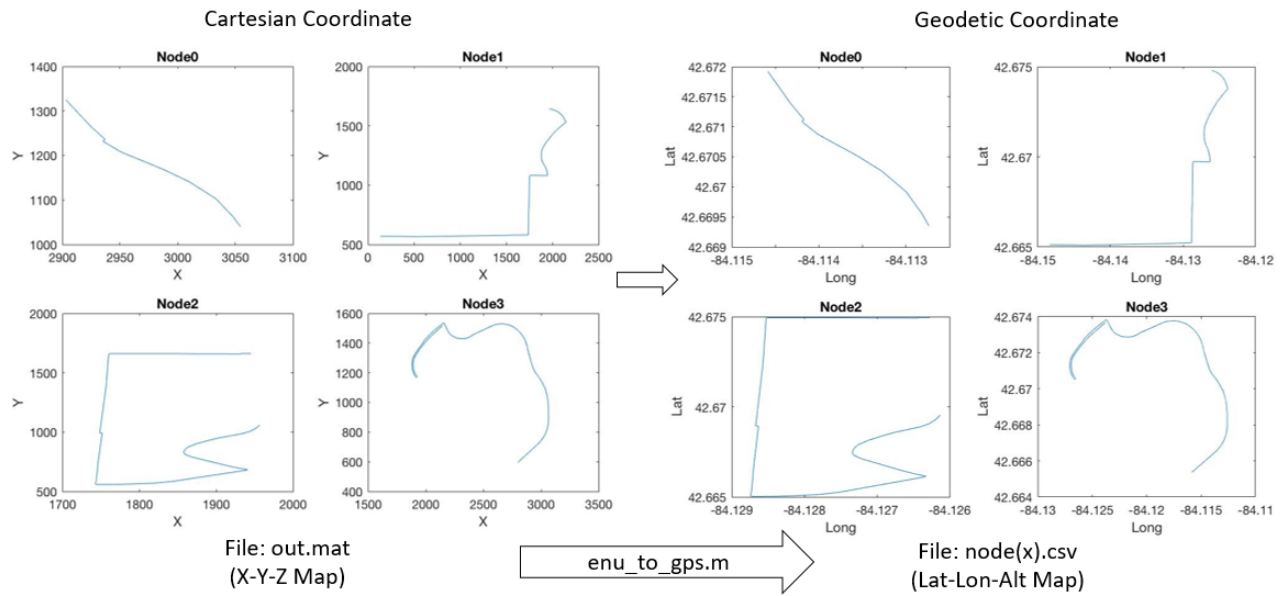

Figure 3-4: ENU to Geodetic Coordinate Conversion

*3.1.2 Remote Vehicle based Log Generator*

The mobility logs generated by the mobility handler is a highly accurate representation of the mobility of vehicles moving around the simulate roads of SUMO generated from the actual maps provided through OpenStreetMap. As mentioned previously, another way to generate the mobility traces is through performing an on-road experiment where the DSRC equipped actual vehicles are used to record their mobility information while moving in a selected testing facility. This method is efficient in a way that the log is more accurate as they are generated through the real onboard sensors of a vehicle in contrast to the SUMO based simulation logs. The basic mobility logs consist of around six primary fields such as Timestamp, Node ID, Positioning through GPS (such as Latitude, Longitude, Elevation), Speed and Heading. These fields provide the most important information required as an input Vehicular Safety Applications. The DSRC devices nowadays are also capable of generating additional fields like path prediction and doesn't require an external application to calculate these fields. Unlike mobility handler, Remote vehicle based traces doesn't require an additional coordinate conversion before replaying the logs. These recorded traces, if required, are then processed and are then separated into multiple trace files based on Node ID. These separated traces are then provided as an input to the CommSim to simulate the on-air CSMA/CA MAC behavior.

The following figure illustrates an example mobility file along with the traces generated using a RV near UCF area.

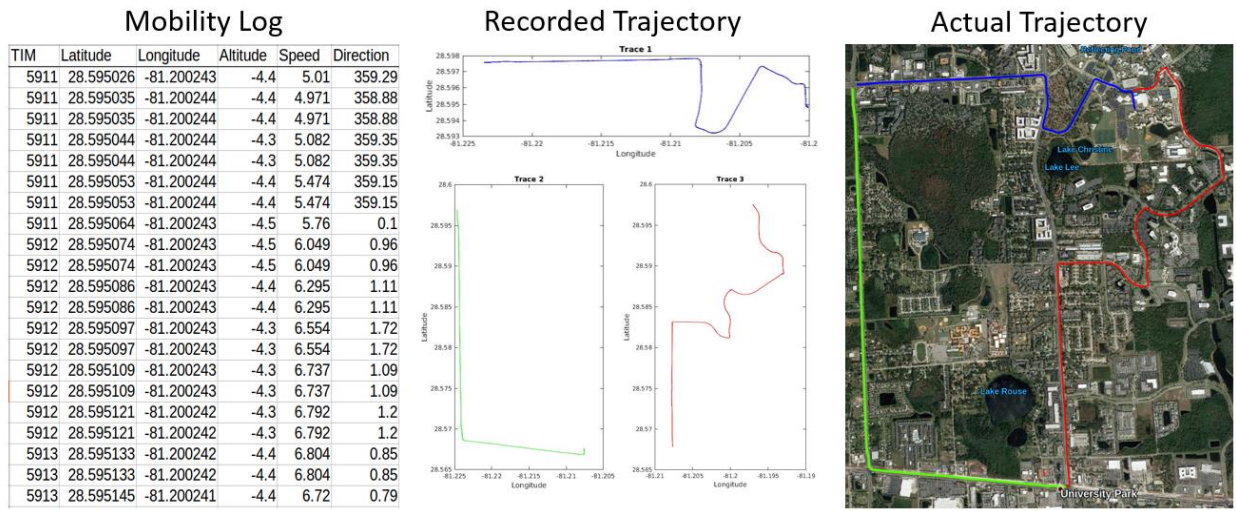

Figure 3-5: Logs generated using Remote Vehicle

## 3.2 Communication Emulator (CommSim)

Communication Emulator or CommSim lies at the heart of our RVE architecture. From the perspective of a single onboard safety unit device, the objective of CommSim is to make the device appear as if thousands of vehicles are transmitting their packets over a DSRC channel. To achieve this, one way would be to use hundreds of OBUs to simulate the behavior of hundreds of vehicles.

Then take the output of this simulator and emulate their collective impact on the channel. However, this method is restricted to the number of OBUs that could be used. In our proposed method, we show that behavior of up to 5000 vehicles (or nodes) can be simulated in real-time using a single device through the communication emulator module.

As mentioned earlier, mobility logs generated from the log generator unit are passed to the CommSim unit. CommSim's task is then to emulate realistic and real-time behavior of MAC and PHY layer for N nodes. DSRC is based on IEEE's 802.11p standard [1]. IEEE 802.11p uses the IEEE 802.11e standard's Medium Access Control protocol in Ad-Hoc mode, which is based on the Enhanced Distributed Channel Access (EDCA) mechanism [15]. The current implementation is based on the pure CSMA/CA protocol in which all nodes are fully connected. The CSMA/CA MAC behavior can be accurately modeled as a Markov Chain where each node performs backoff procedure whenever it senses the channel busy. During the backoff period, node countdowns a random number of idle slots before attempting to transmit again. As soon as a node senses the channel as idle, it waits for a short interval of time called Arbitration Inter-Frame Space (or AIFS, detailed in the next section). At the end of AIFS, if the node finds the channel to be idle again, it transmits the packet immediately. As discussed in section 2.2.1.3.1, there is always a small Propagation Delay (PD) association between any two nodes in the channel. The PD is the delay in sensing a transmission from a remote vehicle node by a receiver vehicle node. Due to this delay, the channel might appear to be idle when a remote node has already started transmitting or even busy when the remote node has completed its transmission. As for all our simulations we will be considering a fully connected network, thus it is safe to assume a small fixed average propagation delay of 5 µs. Hence the design of CommSim should be such to comprehensively enact the MAC behavior of each of the N nodes.

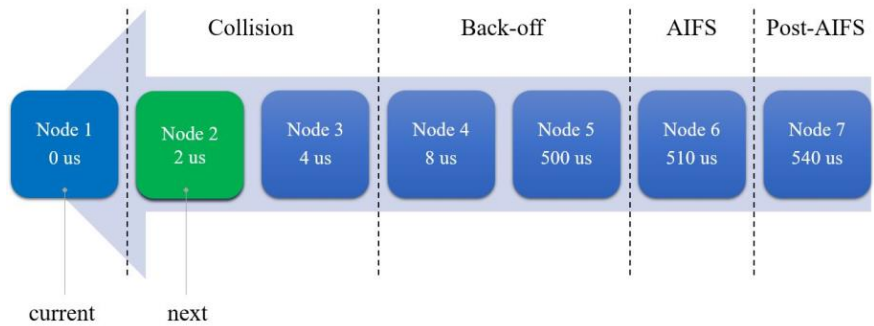

Figure 3-6: The Sorted Nodes Queue

CommSim stores a single packet from each of these nodes into a binary tree data structure known as min-heap queue. The nodes in this queue are always stored in a nearly-sorted order based on their respective scheduled time. Thus, the front element of the queue always holds the earliest scheduled packet. Figure 3-6 shows the representation of the queue with seven nodes. The front element of the queue is the *current packet* while the second element always holds the *next* scheduled packet in the queue. Based on the CSMA/CA MAC behavior, a decision is made for the *next* scheduled packet. Depending upon the scheduled times of these nodes, they fall into one of the following four cases:

### 3.2.1   Case 1: Packet in collision

As discussed in section 2.2.1.1, signal propagation delay has a significant impact on the transmission of current packet. Collision scenario mainly focuses on identifying all the nodes which are trying to transmit within a fixed propagation delay. CommSim iteratively finds all such *next* packets and declares the *current* packet in collision. This is followed by the transmission of an extended packet. This effect contributes to the channel utilization and increases the PER. Figure 3-7 shows next four scheduled packets at 0, 2, 4 and 8 micros respectively. CommSim picks the first packet as a *current* node and then begins iterating through the next scheduled nodes. As the next two packets arrive within the propagation delay (5 micros in this case), CommSim declares all three nodes in collision and sends an extended damaged packet over the channel.

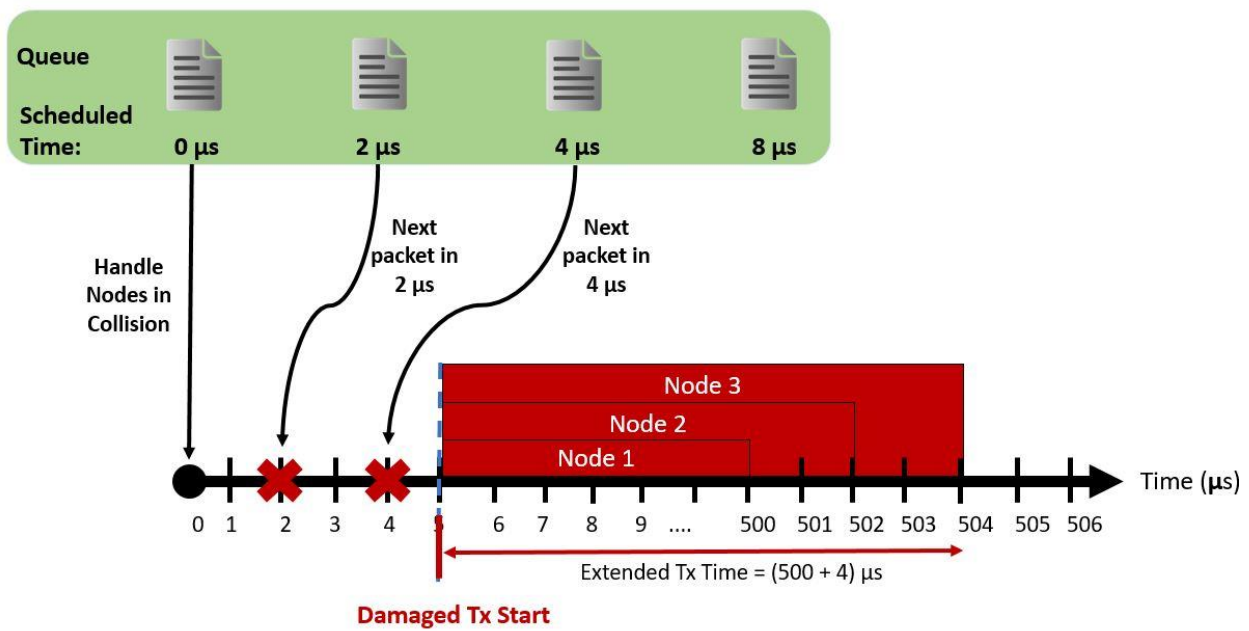

Figure 3-7: Collision Condition

### 3.2.2  Case 2: Perform Back-off

In CSMA/CA, the nodes sense the channel as busy or idle before transmitting a packet. If the channel is busy, the node performs a back-off mechanism. To do this, the node picks a random integer from a uniform distribution between [0, CW], where CW is the Contention Window. As soon as the channel is sensed as idle, the back-off node first waits for an Arbitrary Inter-Frame Space (AIFS) Interval. After AIFS, if the channel is still sensed as idle, then the back-off node starts decrementing its back-off counter with each time slot. If the channel again becomes busy, the node freezes its count till the next AIFS and repeats the procedure until the back-off counter reaches zero. A back-off counter of zero means that the node can start transmitting as soon as it senses the channel as idle and waits for the respective AIFS interval. This behavior is roughly demonstrated in Figure 3-8. After the previous collision scenario, the current node had started transmission. Now the next two nodes in the queue are scheduled at 8 micros and 500 micros. These two next nodes will find the channel as busy and thus perform the above mentioned back-off operation.

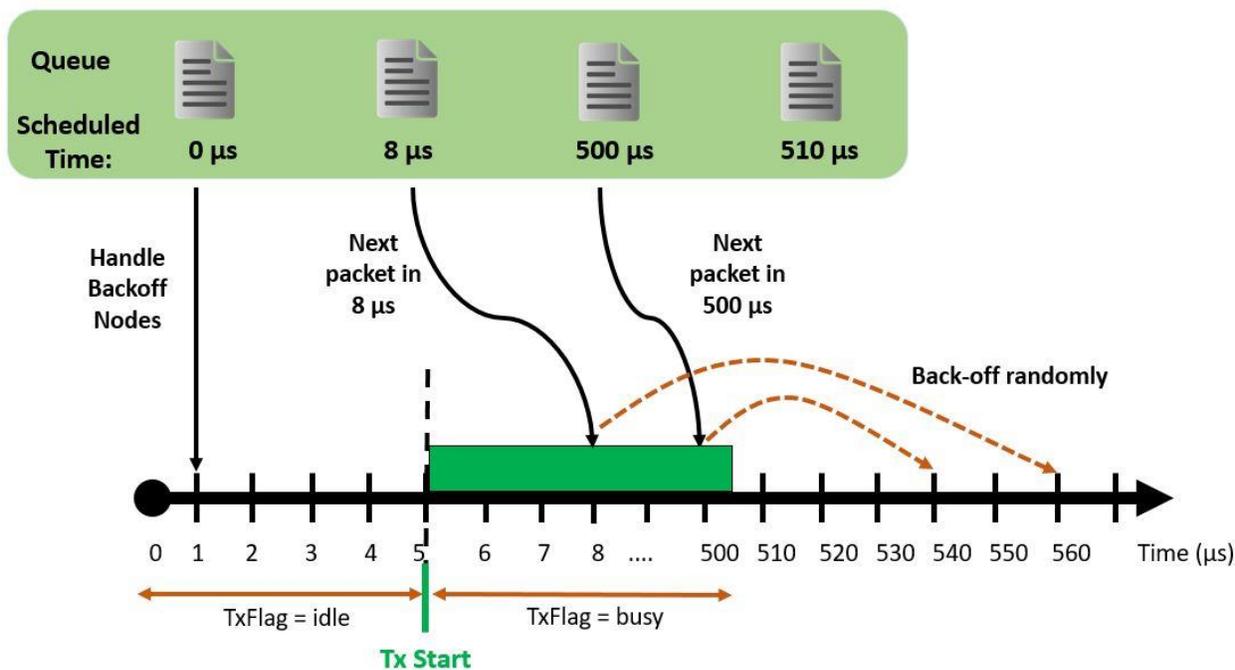

Figure 3-8: Back-off Condition

### 3.2.3  Case 3: Arrived at AIFS

Each transmission in broadcast mode of CSMA/CA is followed by an AIFS time interval. The AIFS ensures the reception of the last transmitted packet by every node in the network range. Each AIFS interval also accounts for one decrement from each back-off node. This prevents any back-off node from starving and guarantees a finite amount of time for acquiring the medium. There is a possibility that a node is scheduled to arrive during the AIFS interval. If this node doesn't have a back-off history, it is not eligible for the back-off mechanism as the time when it arrived, the channel was idle. Instead CSMA/CA reschedules this packet from its current scheduled time to an additional AIFS interval. This behavior is demonstrated by the node scheduled at 510 micros in Figure 3-9. If this node still finds the channel idle, it is eligible to transmit its packet.

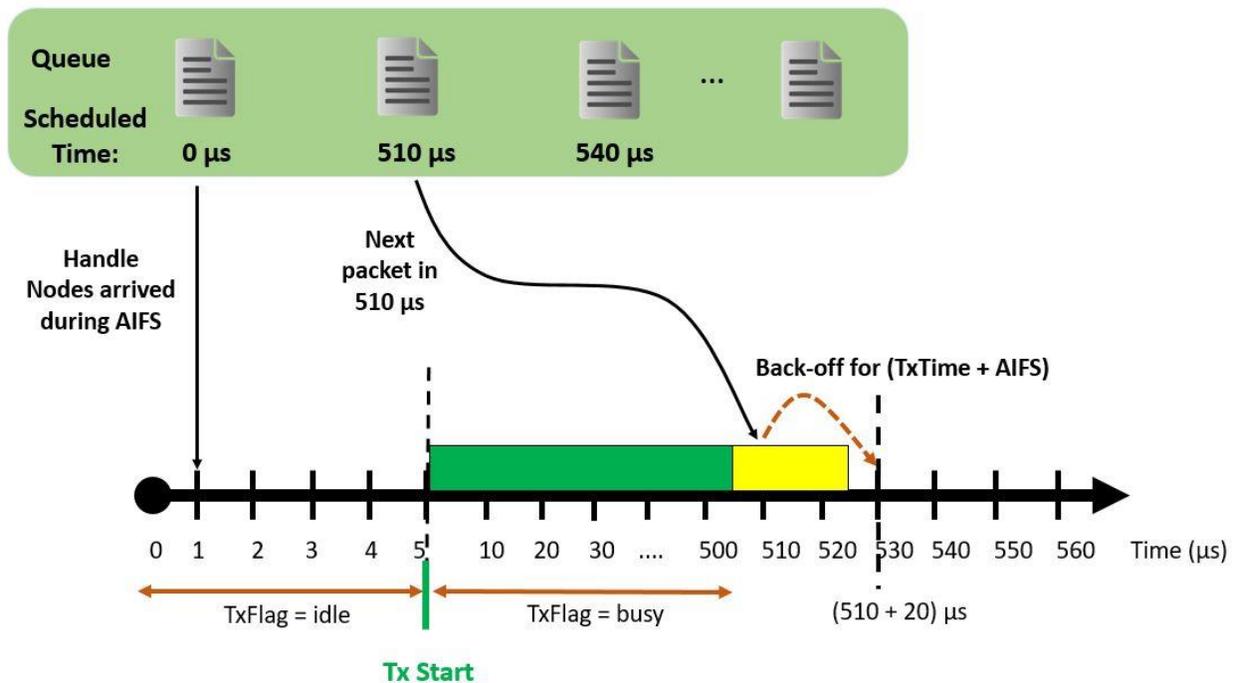

Figure 3-9: AIFS Condition

### 3.2.4   Case 4: No action (Arrived after current transmission or Post AIFS)

The nodes that arrive after AIFS has elapsed are considered in the next transmission as they do not interfere with the previous cycle. In Figure 3-10, node arriving at 540 micros will be ignored in the current transmission.

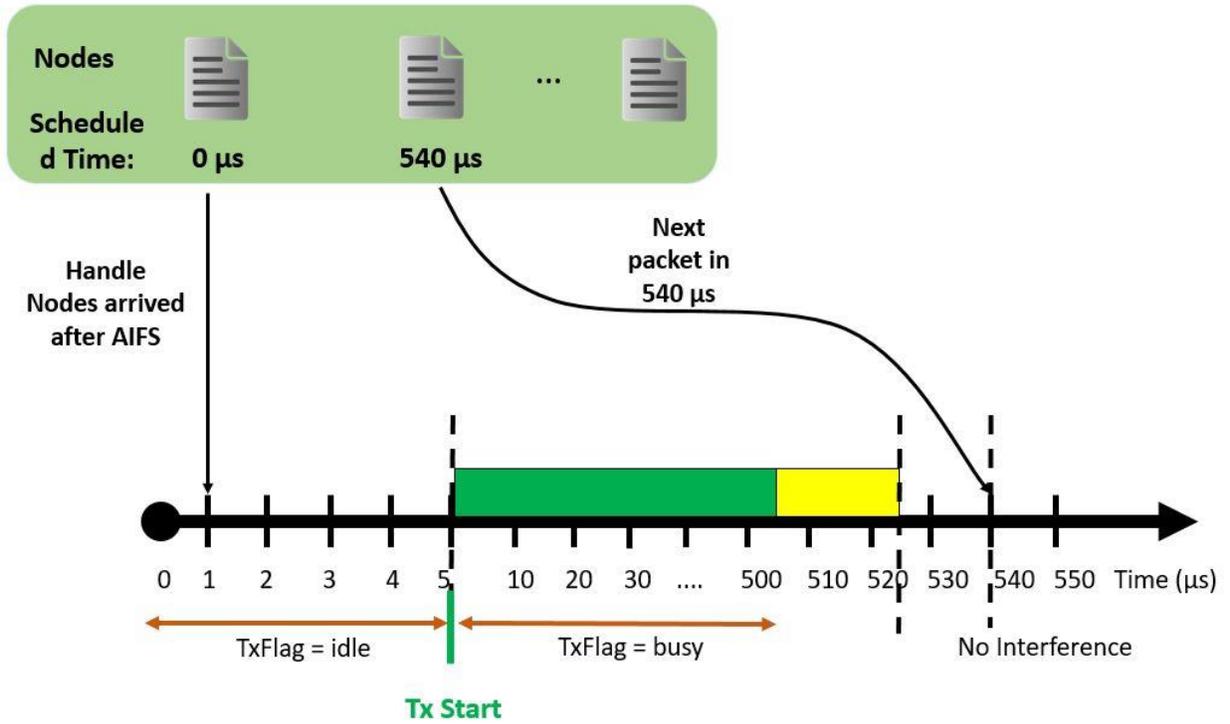

Figure 3-10: Post-AIFS Condition

# CHAPTER 4: IMPLEMENTATION

## 4.1 Mobility Handler

### *4.1.1 Directory Structure*

The RVE hardware mainly consists of two parts, a Linux Host system and a target DSRC device. The Mobility log generation is performed on host system while the actual emulation of these logs is carried out on the target DSRC device. It is highly recommended to follow the original directory structure of RVE.

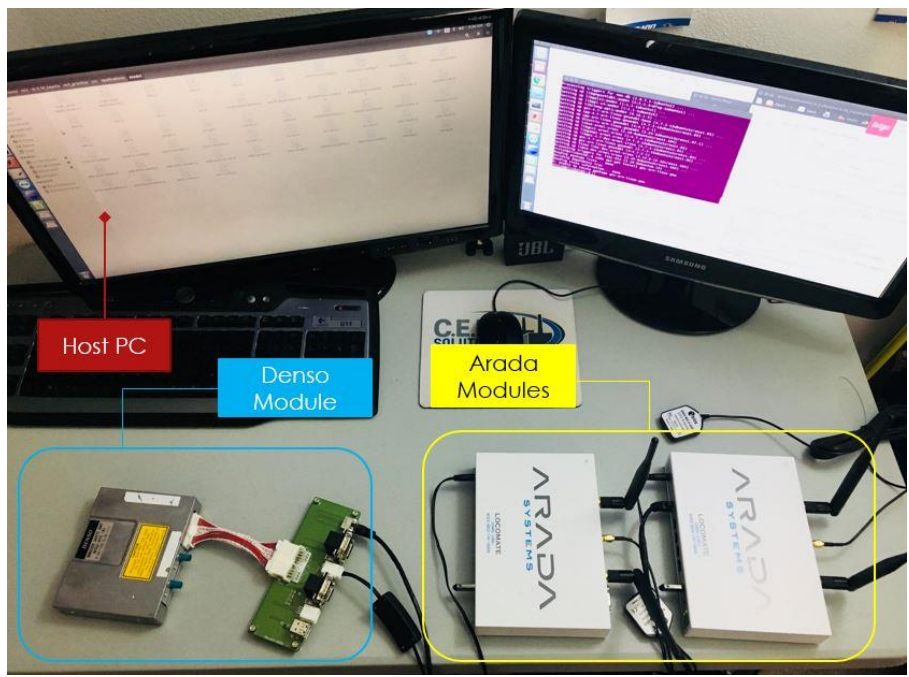

Figure 4-1: Emulator Desk Setup

### *4.1.2 SUMO Configuration*

In the host Linux system, the SUMO must be installed in home/user/ directory. Set the SUMO_HOME environment variable to the installed sumo path in the Bash file. Move the SUMO generation scripts in the SUMO_HOME/scratch/ directory. Create an empty directory in the SUMO_HOME/scratch/ and name it after your test scenario. The next step is to select your scenario in OpenStreetMaps and download an osm file in the newly created scenario directory with the same name as the test scenario. The following directory structure will be SUMO_HOME/scratch/test1/test1.osm. From the scratch directory, execute the generate script with the two arguments: test scenario name (test1 in this case) and the number of nodes (or Vehicles) to simulate.

Example: *sh generate.sh test1 500*

As discussed in section 3.1, this will execute a series of intermediate steps to generate an NS2 mobility trace file. A directory named trace containing the NS2 trace file will be created in SUMO_HOME/scratch/test1/. Execute the trace-to-mat python script inside this directory to create a MAT file for the next step. As discussed, the GPS coordinates in a NS2 mobility trace file are in the cartesian coordinate system. SUMO trace generation process logs the mapping of the origin in the geodetic coordinate to the cartesian coordinate. This mapping can be found in the [logs] file. Using the enu-to-geo MATLAB script, the Cartesian/ENU coordinates are mapped to the Geodetic coordinate system. In this script the reference Lat-Lon has be modified to obtain equivalent geodetic conversion. The output of this script is stored in the SUMO_HOME/scratch/test1/traces/nodes directory as separate node files. Each of these node files consists of fields required such as Node ID, Time, Latitude, Longitude, Heading, Speed and so on. These fields are required for information sharing between vehicles through Basic Safety Messages.

## 4.2 CommSim

At core, CommSim packet encapsulates three objects - BSM, Collision flag and a Backoff counter. Collision flag represents if a packet has collided. Backoff counter keeps a track of random backoff count in the backoff process. Collision flag and Backoff counter of every packet are initialized with 0 and -1 respectively. To emulate N vehicles (or nodes), a queue of size N is initialized with one packet from each node. This packet queue is sorted based on their scheduled transmission time such that earliest scheduled packet is always at the front. Every new packet inserted into the queue takes its position based on the sorting order.

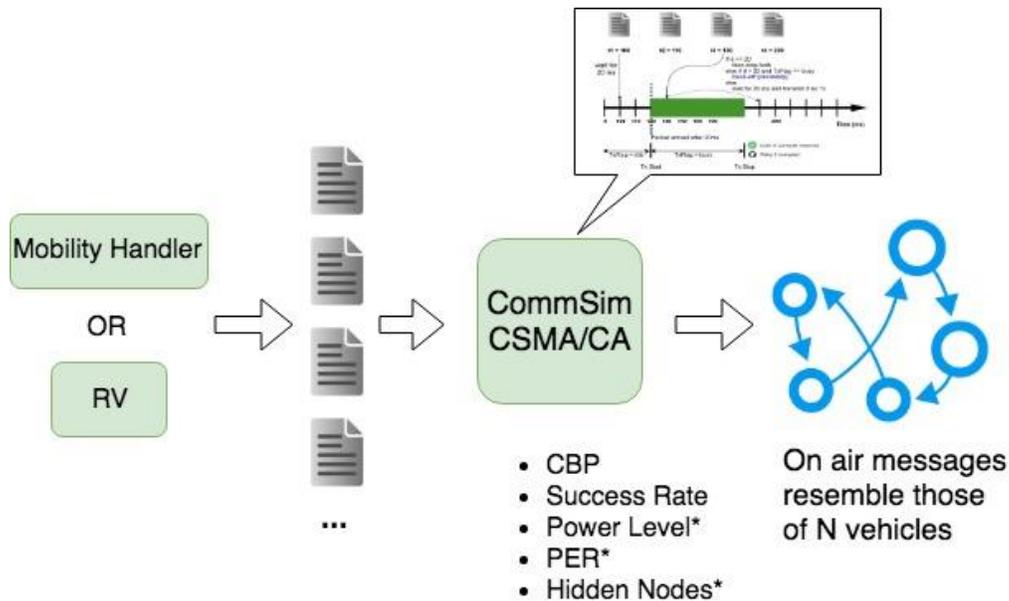

Figure 4-2: CommSim Overview

Figure 4-3 represents a complete flowchart of the CommSim application execution flow. Each time the channel is idle, and a packet is scheduled for transmission, it is popped out of the queue and considered for transmission (referred as curr in Figure 4-3). The packet curr goes

through a look-ahead tree, over the next scheduled packet from the queue iteratively. The tree branch is selected based on the difference (diff) between curr and next scheduled node from the queue. However, if the queue gets empty, the emulator transmits all the remaining packets from the curr node and terminates simulation. Let's consider each of the four look-ahead branches sequentially.

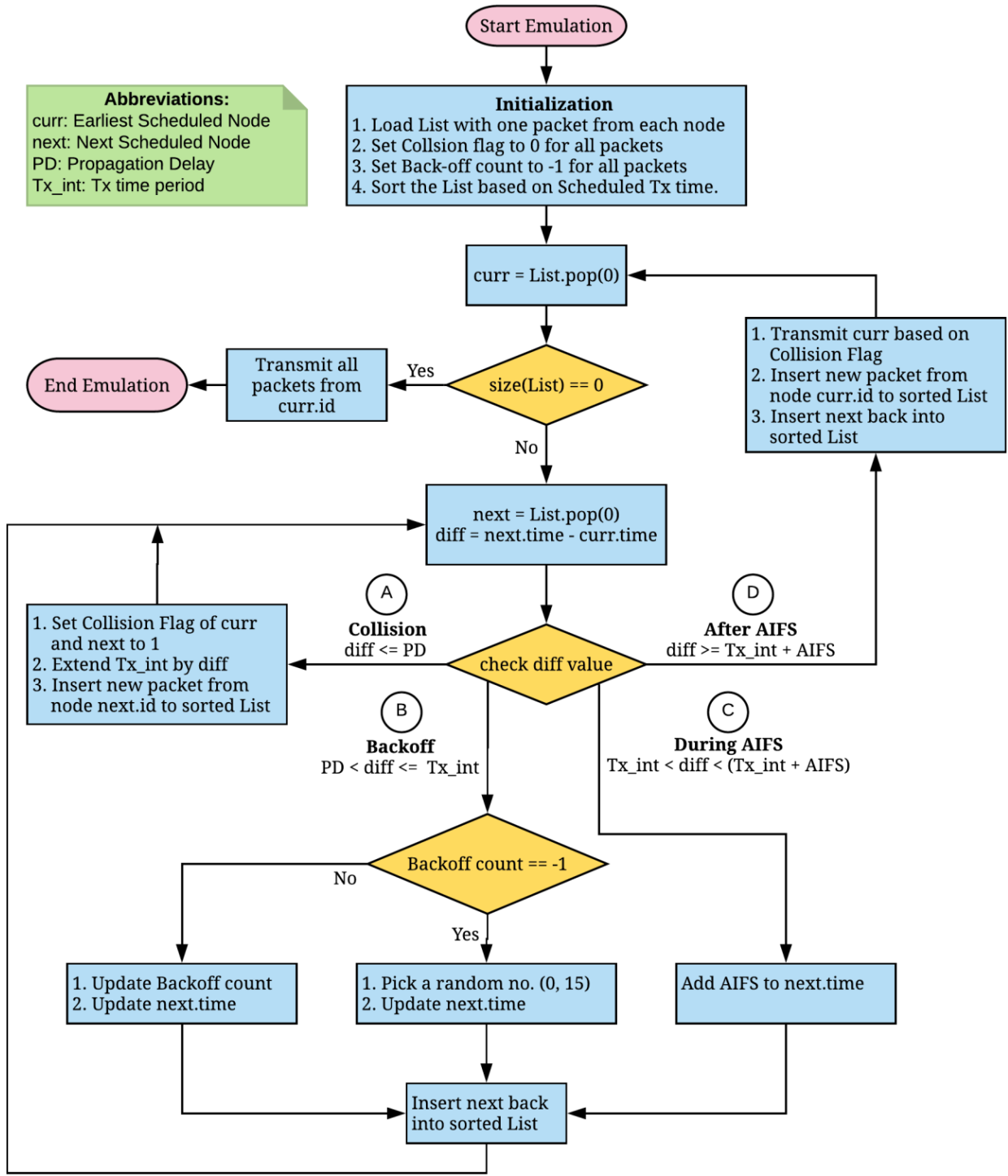

Figure 4-3: Communication Emulator Flowchart

### 4.2.1 Branch Operation

#### 4.2.1.1 Collision

There is always a small Propagation Delay (PD) associated with transmission from one node to the reception by other. A Collision is encountered whenever any two nodes start their respective transmission within the PD interval. Due to the collision, the channel appears busy for an extended period.

In CommSim, this behavior is simulated by the collision branch A as shown in Figure 2. Whenever diff, which is the difference between curr and next node is less than or equal to the PD, a collision is declared. As a result, collision flags of both the nodes are set to 1. In addition, the transmission duration of the damaged packet is extended by diff. Finally, a new packet from the same node represented by next is inserted into the sorted queue. The fact that any two packets from the same node are separated by at least 100 milliseconds prevents the new packet from interfering with curr transmission. Once a next packet with scheduled transmission time greater than PD is encountered, no more collision will occur and the next packet will fall in one of the remaining three branches i.e. Backoff, AIFS, or Post-AIFS.

#### 4.2.1.2 Back-off

If a node has a packet to transmit but it senses the channel as busy, it is required to backoff with a random number. This essentially means that the packet must be rescheduled to a later time defined by random [0, CW] * slot_time. Each time the channel becomes idle, an equivalent number of slots are reduced from the backoff counter of each backoff node.

As shown in Figure 2, the program enters backoff branch when the next packet arrives after PD but before the curr packet transmission ends. Based on whether next packet has a backoff history, two cases may arise. If it had no backoff history i.e. backoff counter is -1, the node is required to pick a random number from 0 to 15. Based on the random number, the transmission time is updated and the packet is inserted back into the sorted queue. However, if the packet has a backoff history, the counter value is updated based on the number of idle slots observed after the previous transmission. Depending on the updated counter value, the packet's transmission time is updated and then inserted back into the sorted queue. This ensures that all the backoff packets are rescheduled and updated in the queue to arrive at a later point in time.

### 4.2.1.3 AIFS

AIFS interval ensures that after each transmission the channel is kept idle for a short interval of time defined by,

$$AIFS = SIFS + 2 * slot\_time \qquad (4\text{-}1)$$

*where*, SIFS (in microseconds) is Short Inter-Frame Space.

AIFS prevents the backed off nodes from starving to transmit in the case when all nodes are lined up one after another (or synchronized). Hence after each transmission, at least one idle slot will be observed by all nodes.

In our implementation, the next packet enters the AIFS branch if its scheduled transmission time is after the *curr* transmission ends but before the AIFS interval ends. The packets scheduled during this duration, are rescheduled to an AIFS interval from their own transmission time. All such next rescheduled packets are inserted back into the sorted queue.

4.2.1.4 Post-AIFS

By this point of time, one of the above three actions, Collision, Backoff, or AIFS has been taken for each next packet that interfered with curr transmission and AIFS interval. A damaged or undamaged packet is transmitted into the channel based on the collision flag of curr packet. The next packet that arrives after the AIFS interval, is considered for transmission (as curr) in the next iteration of the look-ahead tree. The emulation ends when all packets from each node are transmitted.

*4.2.2 Receiver Model*

As discussed in section 2.3, the propagation of a signal through a medium suffers three types of channel impediments, path loss, shadowing and/or multipath fading. Since propagation path loss constitutes to a large-scale signal power loss in a channel, it is essential to consider its effect in emulation. In CommSim the frame is initially set for transmission at 20 dBm power level. However, around 3 dBm power loss is due to cable, and the remaining 17 dBm is the actual signal transmission power. As the signal is attenuated due to path loss in the channel, a nearly deterministic part of the power is lost during propagation through the channel. Thus, the power received at the receiver known as Received Signal Strength Indicator (RSSI), is a function of distance between the transmitter and the receiver.

As the signal power loss is a function of distance, the nodes which are farther from the transmitter experience a larger attenuation than the ones that are closer. A signal can be successfully decoded from the received noisy signal if it's Signal Interference to Noise Ratio (SINR) is greater than a set threshold. The SINR of a received signal is defined as the ratio of the power of a certain signal to the total power of the received interfered signal. Thus, SINR can be expressed as,

$$SINR = \frac{S}{I+N} \tag{4-2}$$

$where, S: Power\ of\ measured\ usable\ signal$

$I: Interference\ Power\ of\ other\ signals$

$N: Background\ Noise$

Since in CommSim, the distance of all frame participating in collision is known ahead of time, it is possible to simulate receiver model by only sending the packet whose SINR is calculated to be greater than the SINR threshold. For instance, if three other packets arrive within 5 µs of a scheduled packet, the collision is declared and a damaged packet was to be sent. Instead we utilize the knowledge of distance between nodes and calculate the RSSI values for each of the four packets before sending a garbage packet. Based on the calculated RSSI value, SINR values can be calculated for each of the four signals. If any of the signal has a SINR value greater than the SINR threshold, it is sent over the channel without any interference of other three nodes. Presence of a receiver model in a simulator is essential as it can accurately model the losses in the channel due to impediments as well as other participating nodes.

### 4.2.3 Multi-Threaded Transmission

The CommSim uses a Min-heap binary tree data structure to store the scheduled packets from different node into a queue. As discussed in section 4.2.1, during communication simulation, the current and next nodes are iteratively fetched (deleted) from or inserted into this queue several times during a single packet transmission. The time-complexity for insertion or deletion from a Min-heap based queue is O(log n) which is discussed in Appendix A in a greater detail. However, even though the insertion or deletion aren't time consuming operations, a very small fraction of additional simulation time is expended in performing these operations inside loop. This time difference is not visible until the number of nodes exceeds the threshold of 5000 nodes. Even though it is highly unlikely to consider nodes more than 5000 for testing vehicular safety applications, this problem can be easily mitigated by careful design of emulator application. The idea behind solving this problem is the fact that the transmission time of a packet from a node takes much higher time than going through each eligible next node and branching through the application iteratively. From the implementation point of view, the transmission of a packet from a node can be independently spawned as a separate thread after going through each of the four conditions of the CommSim. While this thread transmits a packet in the medium, the main application has enough time to handle each of the next packet in the queue for the next cycle. This utilizes the transmission time taken by the current node, for making decision about the next scheduled packets ahead of their turn. This makes the CommSim capable of handling even larger number of nodes efficiently and in real-time. Thus, all the computation takes place parallelly during a small fraction of transmission time and doesn't affect the actual simulation time.

### 4.2.4   *Logging in CommSim*

CommSim provides a feature to log all the packets based on the decisions made during the communication simulation. This includes details such as the next scheduled packets from different nodes in the queue, action taken on the current packet and on each of the next scheduled packets iteratively. In addition, it also logs the status of the last transmitted packet and few channel details such as CBR and PER. This is immensely helpful when analyzing the behavior of any node in the simulation or during debugging. The generation of log can be disabled from the application. Figure 4-4 shows a sample CommSim generated log.

Figure 4-4: CommSim Logs

### *4.2.5 CommSim Configuration*

Firstly, the folder consisting of separate node traces must be transferred to the target DSRC device. The corresponding path to the nodes directory should be provided in the CommSim configuration. The CommSim configuration has several parameters that must be provided before the simulation is carried out. Let us discuss each of these parameters in greater detail.

1. NODES: This parameter specifies the total number of vehicles to be emulated
2. SLOT: This parameter specifies the minimum slot duration in time

3. Scenario: This parameter specifies the path to the separate mobility traces
4. Propagation Delay: This parameter specifies the minimum fixed time difference between transmission and sensing of a signal from a remote node
5. AIFS: This parameter defines a fixed time duration for which the channel is kept idle after each transmission
6. Ptxtime: This parameter specifies a fixed packet transmission duration
7. Sinr_th: This parameter specifies the received SINR threshold for successful reception

## 4.3   Network Simulator-3 Configuration

### *4.3.1   Reverse Scheduling*

The comparison of CommSim with NS3 requires that the same mobility traces are replayed by both simulator. This involves the usage of similar timing and positional data so that the MAC receives the same scheduled packets for each node. Also, the trace positioning data matching is required as the Propagation model and the Congestion control techniques rely on the distance between the nodes. Although NS3 can read the NS2 mobility traces to simulate the communication, it only considers the position data from the traces. Thus, the packet scheduled timing data from the trace is neglected in NS3. Instead, NS3 uses its own virtual timing generator module to generate the packet schedule timings. The NS3 Configuration that we considered generates new packet from each node in every 100 milliseconds. However, with this defaults timing mechanism, all packets from each node are synchronized and would results in higher collision rate as every node attempts to send their packet simultaneously. With an introduction of a small jitter in the scheduled timings this effect can be modified to a much-randomized nodes scheduling. For our experiments, we take both cases into account, synchronized as well as randomized. For randomized simulation in NS3, each time a node is scheduled for transmission, we add a small jitter value generated from a random uniform distribution ranging from -2 to +2 milliseconds. The randomization can be reduced by changing this range by considering a smaller jitter value. For our experimental evaluation we consider the synchronized nodes simulation with a jitter range of ±400 microseconds. On the other hand, for a randomized nodes simulation, we set the jitter range as ±2 microseconds.

Similar packet schedule timings are essential to make both simulator comparable to each other. Since NS3 doesn't allow to provide timings from the trace logs, we took the reverse approach such that, we log all the packets from each node generated by NS3 just before they are handled by the MAC layer into a single trace file. Along with this, NS3 logs the Channel Busy Percentage (CBP) and Packet Error Rate (PER) metrics which are later used to compare with the same metrics logged using CommSim. This reverse approach is taken only for the comparison purpose and doesn't need an external mobility trace generator. The single mobility log is then separated into different nodes log. These nodes are then replayed using CommSim to generate a comparable BSM rich cloud and simulate the MAC behavior in a much similar way as NS3. The only difference is that the same NS3 simulation takes much longer time as compared to the CommSim based real-time simulation.

*4.3.2  Mobility Model*

In general, for the current version of CommSim, any mobility model in which all the nodes are fully connected to each other can be used for simulation. However, for simplicity, we used the following two different types of mobility models provided by default in NS3.

4.3.2.1  Grid-based Model

In a Grid based model, all nodes were placed inside a grid of fixed dimensions. Each node inside a grid was allotted a position at the start of the simulation. The mobility of each node was based on a Constant velocity mobility model provided in the default NS3. Since this network of node is required to be fully connected, we placed all nodes in a grid of width 60 meters with each node placed 2 m apart in both x and y direction as shown in Figure 4-5. This arrangement places 30 nodes in a single line. Thus, if we are simulating 500 nodes, the grid size would be equal to 60 x 34 (~30 nodes on each of the 17 grid lines). Thus, the maximum separation between any two nodes in this mobility model doesn't exceed 68 m which is the diagonal distance of the grid. Since DSRC devices can communicate with each other up to a maximum distance of about 350 m, the above mobility model is a fully connected network and can be used in our simulation for as many as 5000 nodes.

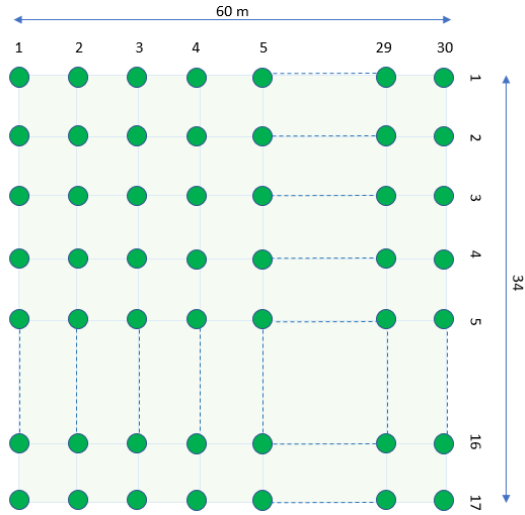

Figure 4-5: Grid Based Mobility Model

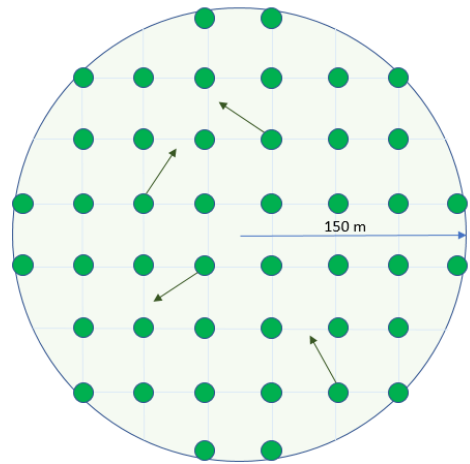

Figure 4-6: Disc Based Mobility Model

4.3.2.2  Uniform Disc based Random Walk Mobility Model

In addition to the grid based constant velocity model evaluation, we performed similar experiments with another kind of position and mobility model provided by NS3. In the model, the nodes were placed uniformly inside a disc shaped position allocator as shown in Figure 4-6. A Random Walk mobility model was employed to make the nodes move inside a bounded rectangle with a randomly assigned constant velocity. The nodes move with a fixed velocity for a certain delay period specified by a delay parameter. After each delay period, the nodes change their direction and speed. This can be modified to trigger based on distance instead of time. For our evaluation, we considered a disc of radius 150 m with all the nodes placed uniformly at the start of simulation. As we used a trigger based on time, the nodes change their speed and direction after every 1 second. Radius of 150 m ensures that all nodes are in a visible range of each other with a maximum distance of 300 m between any two nodes inside the disc.

## 4.4 Emulator Graphical User Interface

The entire process of generating logs using mobility handler and then simulating the communication through CommSim could get cumbersome. To streamline the whole process, a Graphical User Interface (GUI) is designed specifically for RVE. Figure 4-7 shows the basic design of the vehicle emulator GUI. The interface has two main section, Mobility Handler (Left) and Communication Simulator (Right). The Mobility handler interface makes it easy to generate logs by going through each step sequentially. It mainly consists of features like, importing a map from OpenStreetMap, generating the NS2 trace for the selected map based on the number of nodes required to simulate followed by the required conversion between GPS coordinates and finally splitting a single combined trace into separate traces for each node. Similarly, the CommSim interface consists of properties such as the number of nodes to be simulated, the path of previously generated trace files, the simulation time and finally a drop-down to select the type of propagation loss model from a list of several available models. To analyze the behavior of the simulation, an option is provided to generate an additional simulation log file along with its path. Many other features such as saving the current scenario and the selected parameters or loading an existing scenario along with all the parameters would further make the interface easy to repeat the previous experiments.

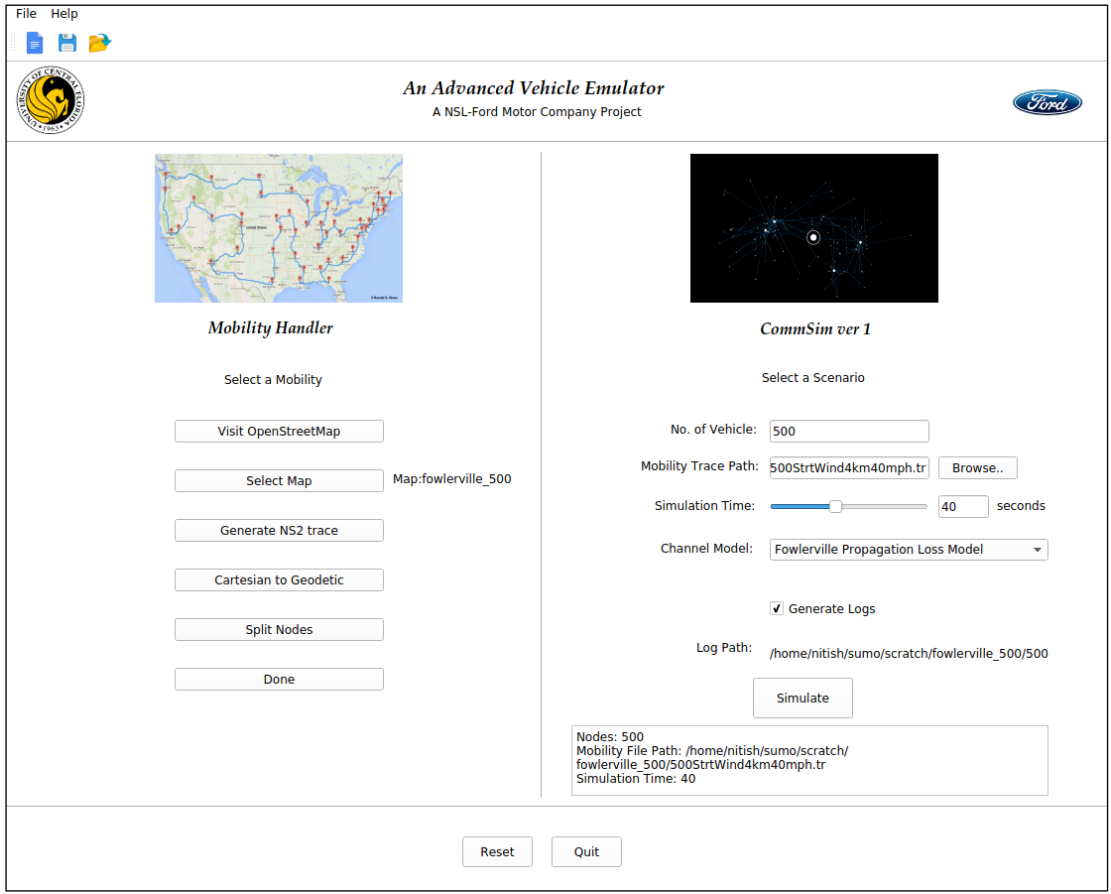

Figure 4-7: RVE GUI

# CHAPTER 5: EXPERIMENT RESULTS

NS3 is a widely used and easily configurable network simulation tool which have been adopted for simulation and testing of CAV applications. In NS3, the underlying simulator operates in dimensionless virtual time. Due the high communication latency and non-real-time performance, NS3 is unsuitable for real-time emulation-based testing. Despite NS3's high-latency, it is a great simulation tool for benchmarking the communication behavior among multiple vehicles. Thus, a comparison between NS3 and CommSim based on simulation results would provide a sufficient validation for the behavior of CommSim. To make NS3 and CommSim comparable, reconfiguration of both tool is essential. The simulation specific parameter settings is summarized in Table 5-1.

Table 5-1: System Parameters used for Simulations

| Parameter | Value |
|---|---|
| AIFSN | 2 |
| Contention Window (min) | 15 |
| Transmission Frequency | 5 Hz |
| SIFS | 32 us |
| Channel Bandwidth | 10 MHz |
| Transmission Power | 20 dBm |
| Slot Time | 13 us |
| Simulation Time | 40 s |
| SINR Threshold | 3 dBm |

The primary metrics that we have considered to analyze the communication channel performance are Channel Busy Ratio, Packet Error Rate and Simulation Time. Let us explore these metrics in more detail.

## 5.1 Evaluation Metrics

### *5.1.1 Channel Busy Ratio*

In cooperative vehicular networks, the adaptive congestion control techniques based on transmission power, packet generation rate or the packet length have been studied. Each of these technique uses a common metric for evaluation depending on the channel load conditions. Thus, each time the channel is assessed based on channel utilization also known as Channel Busy Ratio (CBR). CBR specifies a ratio of the time for which the medium was occupied or busy within a certain time interval. The CBP has been proved to be one of the most suitable metric to improve the performance of a network and maximum the utilization. Few techniques only use the local the CBR through direct sensing while other applications share the CBR information to make more accurate inference about the channel state. In our evaluation, we only consider the local CBR values which are logged every 100 milliseconds.

### *5.1.2 Packet Error Rate*

Packet Error Rate is a measure of the number of received packets in error after the Forward Error Correction (FEC) among the total number of packets that were transmitted. As the CSMA/CA MAC protocol is a best effort based collision handling technique, it inherently holds a probability of collision due to many reasons. If a channel is highly congested and the nodes are nearly synchronized, the channel faces a high volume of collisions. The propagation delay problem is one of the very main source of collision in a relatively dense network. The hidden node effect can also contribute to a large portion of packets that are received in error. Furthermore, PER is highly dependent on the channel properties as the path loss, fading and shadowing could worsen situation. In the case of vehicular safety applications, PER is one of the very critical concern as the increase in PER also increases the tracking error and could lead to several application failures. Careful modeling of the channel and consideration of several congestion control techniques could decrease the packet error rate drastically. We use PER as one of the evaluation metric as it will indicate the performance of the algorithm.

### *5.1.3 Simulation Time*

One of the most important aspect of emulating multiple nodes using a single device is the real-time performance. As the intended application of the proposed RVE design is testing of safety applications, any delay in simulating and sending the messages over the channel is intolerable. As many high-fidelity simulators already exists that can simulate several nodes based on the MAC behavior but none of them, including NS3, gives a real-time simulation of the protocol. Since the CommSim has been designed in such a way that it provides a real-time simulation of the communication between any number of nodes. In a hindsight it uses the most optimum data structure called Min-heap for the behavior model of CSMA/CA protocol. Thus, the required operations can be performed in a limited amount of time given by $O(\log n)$. Even with the increase in the number of nodes to be simulated, the time-complexity of the algorithm remains bounded. Thus, the evaluation of CommSim based on Latency suggests why CommSim should be preferred over other simulators in all time-critical applications especially those involving safety.

### 5.2    Comparison of CommSim and NS3

According to the study in [8], the acceptable range of the CBR values is between 0.4 and 0.8 (optimally around 0.68). It is also shown that when CBR value is above 0.8, the channel experiences high collision rate while a CBR value below 0.4 results in significant underutilization of the channel. As we will see later in this section, the effect of CBR value on PER is clearly visible in our simulation results.

For the comparison of NS3 and CommSim we considered three scenarios where 100, 500 or 1000 vehicles placed inside a square grid of side 160m. The spacing between any two consecutive nodes is 5m, which means that the longest distance between any two nodes is always less than 250m. This ensures that all nodes are in the transmission range of each other. Furthermore, the comparison is performed for both Synchronized as well as unsynchronized transmission behavior. For CBP below 0.68, the synchronized transmission should see more collisions as compared to unsynchronized transmission. This is because the nodes in synchronized case try to transmit nearly at the same time in every 100ms interval.

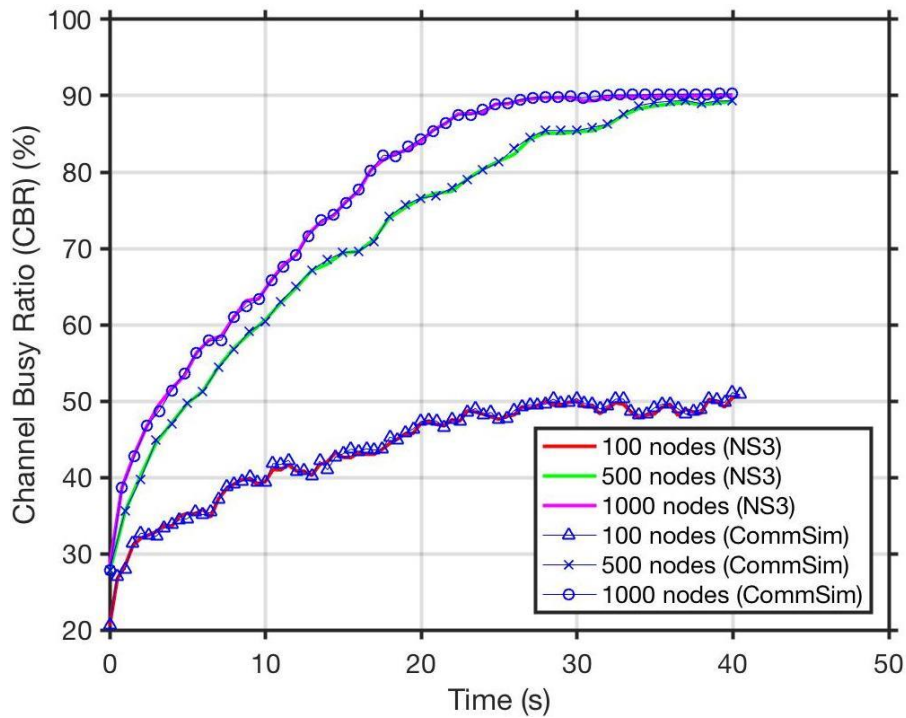

Figure 5-1: CBR Comparison of NS3 and CommSim for Synchronized Transmission

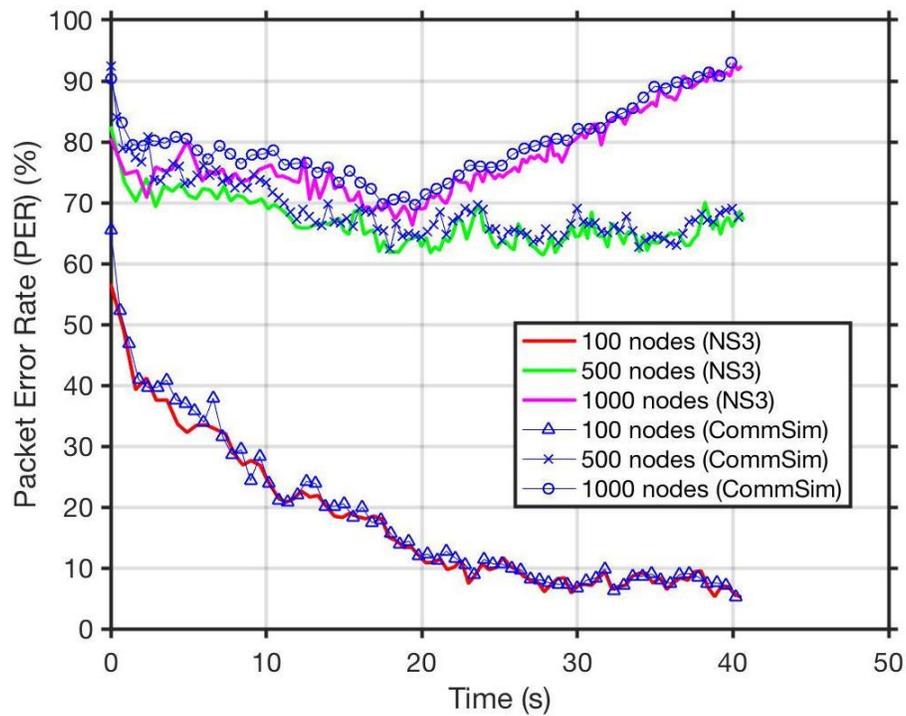

Figure 5-2: PER Comparison of NS3 and CommSim for Synchronized Transmission

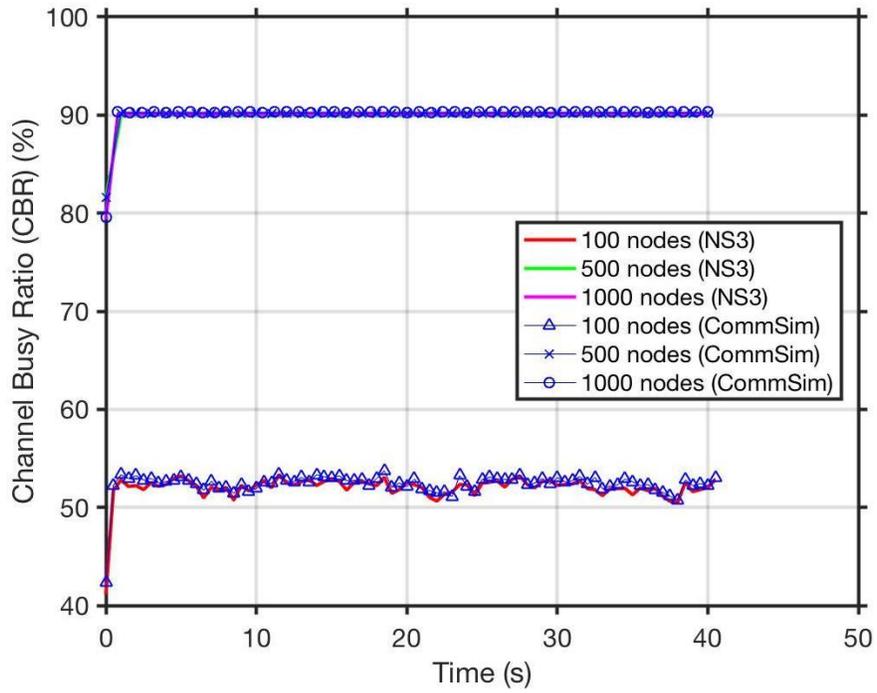

Figure 5-3: CBR Comparison of NS3 and CommSim for Unsynchronized Transmission

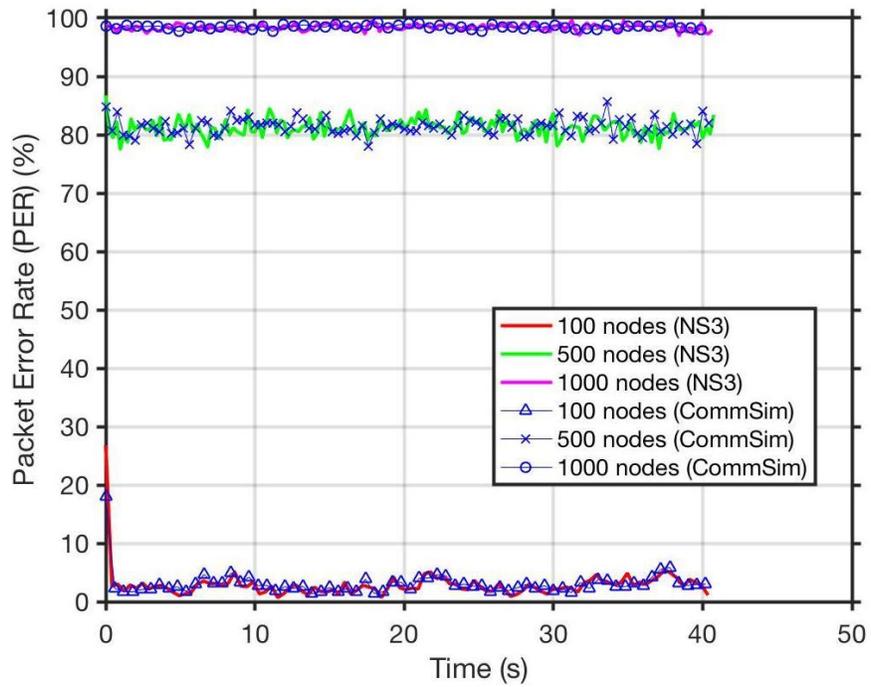

Figure 5-4: PER Comparison of NS3 and CommSim for Unsynchronized Transmission

Figure 5-1 through 5-4 shows a 40 seconds simulation results for synchronized and unsynchronized transmission respectively. Each of these figures shows a comparison between NS3 and CommSim based on CBP and PER. The average CBP of CommSim for all cases lies within ±0.7% of NS3. On the other hand, the average PER of CommSim lies within ±2% of NS3. Table 5-3 summarizes the average CBP and PER for each of the before-mentioned scenario. The slight deviation is essentially due to the randomness in the channel. In Figure 5-2, note that initially, the PER is high due to synchronization in transmission. But since the channel is underutilized, it tries to recover from high PER with time. In case of 100 synchronized nodes in Figure 5-1, the CBP rises to an acceptable range (0.4 < CBP < 0.8) and thus the channel loss recovers significantly over time which can be observed from the decreasing PER plot in Figure 5-2. For 500 and 1000 synchronized nodes, the channel attempts to recover from node synchronization but at around 20th second the channel begins to congest (CBP > 0.8) and thus the PER starts ascending. The delay in the increase of CBP between the 100, 500 and 1000 synchronized nodes scenario is because the synchronization between the nodes breaks easily as the number of participating nodes in transmission increases. Similarly, in Figure 5-4, the 100-node case constantly remains in the acceptable CBR range and thus suffers minimum packet loss. While the channel with 500 and 1000 nodes is congested (CBR > 0.8) from the beginning and thus suffers from high PER.

The premise over which CommSim proves to be a powerful tool for emulation is its low-latency performance. Table 5-2 shows a comparison of simulation time of above experiments in NS3 and CommSim. In case of 5000 nodes, a 40 seconds simulation took as long as ~32 minutes in NS3, whereas the same simulation took 41.99 seconds in CommSim. The additional 1.99 seconds is expended in transmitting the backoff nodes towards the end of the simulation. For each scenario, the simulation time in NS3 is considerably higher than CommSim. Thus, CommSim not only models the communication behavior of N nodes accurately, but also have minimum latency as compared to NS3.

Table 5-2: Simulation Time Comparison

| Number of Vehicles | NS3 (*minutes*) | CommSim (*minutes*) |
|---|---|---|
| 100 | 0.51 | 0.4 |
| 500 | 4.97 | 0.41 |
| 1000 | 12.94 | 0.41 |
| 5000 | 32.03 | 0.42 |

Table 5-3: Comparison of Mean Metric Value

| | No. of Nodes | CommSim | | NS3 | |
|---|---|---|---|---|---|
| | | CBP | PER | CBP | PER |
| Sync | 100 | 43.84 | 81.31 | 43.65 | 83.96 |
| | 500 | 71.47 | 31.12 | 71.33 | 33.93 |
| | 1000 | 76.35 | 20.28 | 76.31 | 21.99 |
| Unsync | 100 | 52.37 | 96.96 | 51.99 | 97.11 |
| | 500 | 89.96 | 18.55 | 89.75 | 16.78 |
| | 1000 | 90.08 | 1.53 | 89.87 | 1.6 |

## 5.3 Effect of Channel Loss Model

As discussed in section 4.2.2, the inclusion of a receiver model is essential to report a realistic value of PER at a given instance in a communication channel. We have already seen the comparison of CommSim with NS3 for the communication simulation without a receiver model. In a communication model the RSSI value of the transmitted signal can be easily approximated by using a suitable channel loss model provided the distance between the transmitter and receiver is known. The latest version of CommSim can effectively simulate the behavior of receiver model by calculating the RSSI values for each node colliding node based on a selected channel loss model. Thus, the signal with the SINR above certain fixed threshold is the only signal which gets transmitted in the channel successfully instead of just declaring a collision as in the previous versions. Figure 5-5 shows SINR as a function PER. Thus, higher the signal SINR, greater is probability of successful transmission.

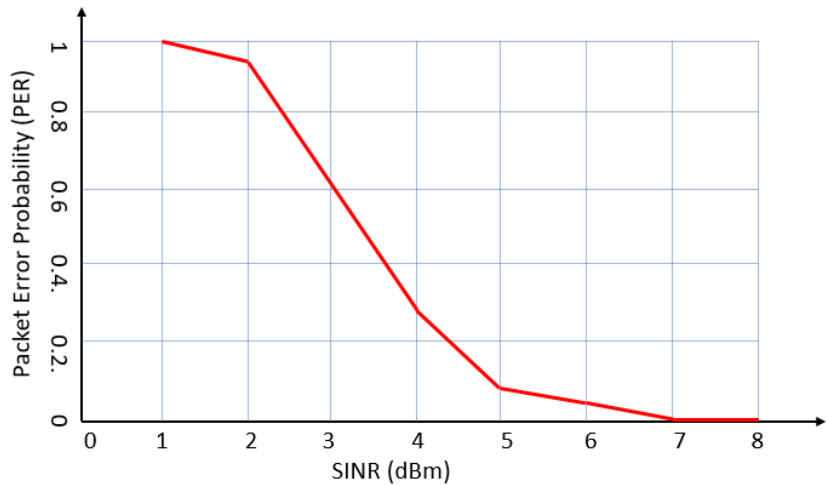

Figure 5-5: Relationship between SINR and PER

The Figure 5-6 and 5-7, shows the effect of using a receiver model on the PER in CommSim for both synchronized and unsynchronized case. From Figure 5-6, it can be inferred that the PER for 100 nodes scenario is approximately reduced by a factor of 20% while the 500 and 1000 nodes scenario experience a reduction of approximately 15% and 17% respectively. Similarly, in Figure 5-7, the PER for 500 and 1000 nodes scenario is reduced by approximately 14% and 18% respectively. Since in this case the 100 nodes scenario was already performing at nearly 0% packet loss, the difference of adding loss model cannot be observed. Thus, CommSim efficiently models the channel loss and receiver model resulting in considerable increase in packet success rate.

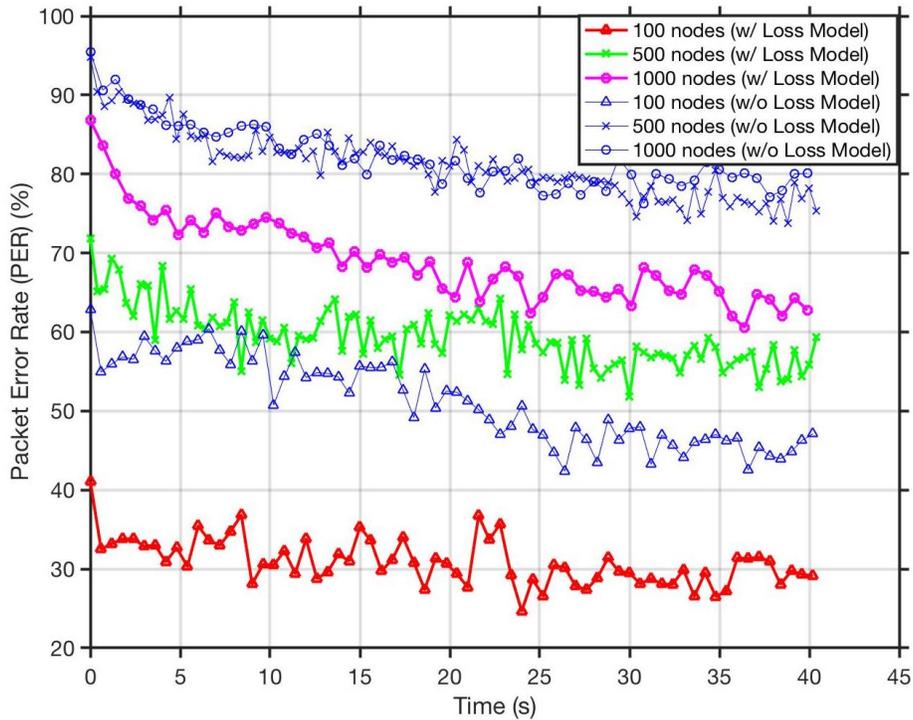

Figure 5-6: Effect of Loss Model on PER for Synchronized Nodes

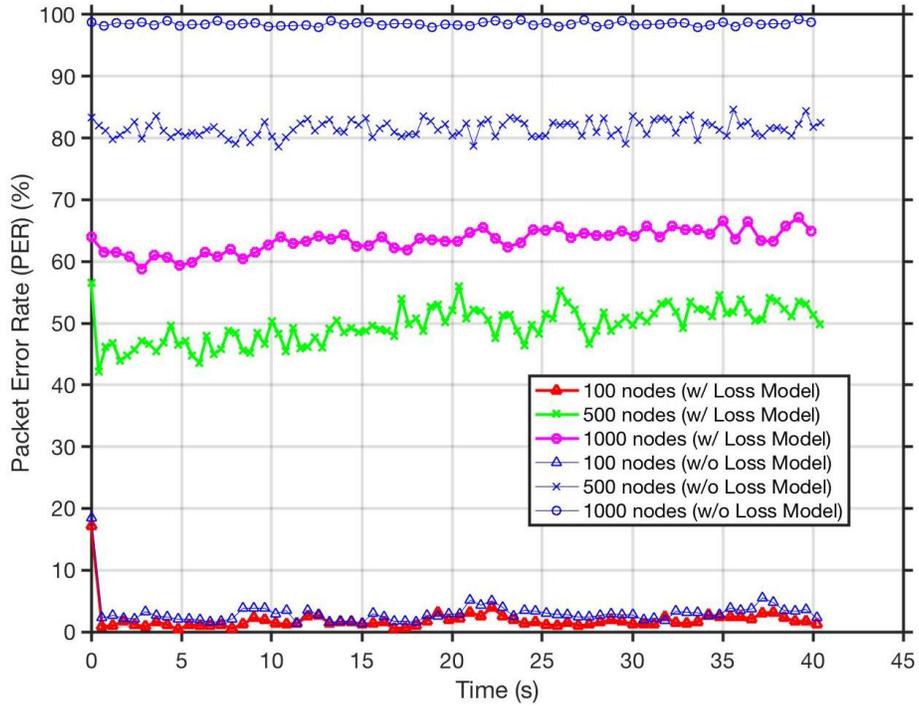

Figure 5-7: Effect of Loss Model on PER for Unsynchronized Nodes

# CHAPTER 6: CONCLUSIONS AND FUTURE WORK

In this work, we have presented a unique, affordable, and easily configurable Remote Vehicle Emulator framework for testing of ADAS safety applications. We then discussed two main components of the RVE architecture, namely Mobility Handler and Communication Emulator. As we have seen, Mobility Handler is independently responsible for creating mobility traces from either an actual vehicle or using a traffic simulator like SUMO. Later, these trajectory traces were fed to a Communication Emulator application called CommSim. CommSim is responsible for simulating the MAC behavior of each emulated vehicle. The channel simulation covers all the subtleties of CSMA/CA to make the emulation realistic and real-time. Performance of CommSim is validated by comparing the simulation results with a well-known simulation tool - NS3. Finally, the evaluation shows that the channel CBR and PER have nearly an identical trend in both cases but with a low communication latency for CommSim than compared to NS3.

With the aim to improve safety of Cooperative Automated Vehicles, the proposed RVE architecture provides a robust ADAS testing tool. This architecture can be enhanced by including a mathematical model as proposed in [8], to mitigate the effect of hidden nodes on channel performance. It can be further extended to improve the channel performance using the distance dependent congestion control strategies based on parameters like power, rate and/or message content as proposed in [9], [11], [12] and [16].

As a realistic traffic condition suffers heavily due to the Hidden Node problem discussed in section 2.2.1.2, the future version of RVE would be targeted towards the inclusion of a mathematical model to approximate the effect due to hidden nodes in a congested and partially observable network. This would help us to predict and reduce the collisions occurred due to hidden nodes. Additionally, a probabilistic transmission mechanism based on various control parameters such as data rate, transmission range, vehicle density, CW etc. can be incorporated to enhance the quality of simulation further. Figure 6-1 provides a reference model for this task.

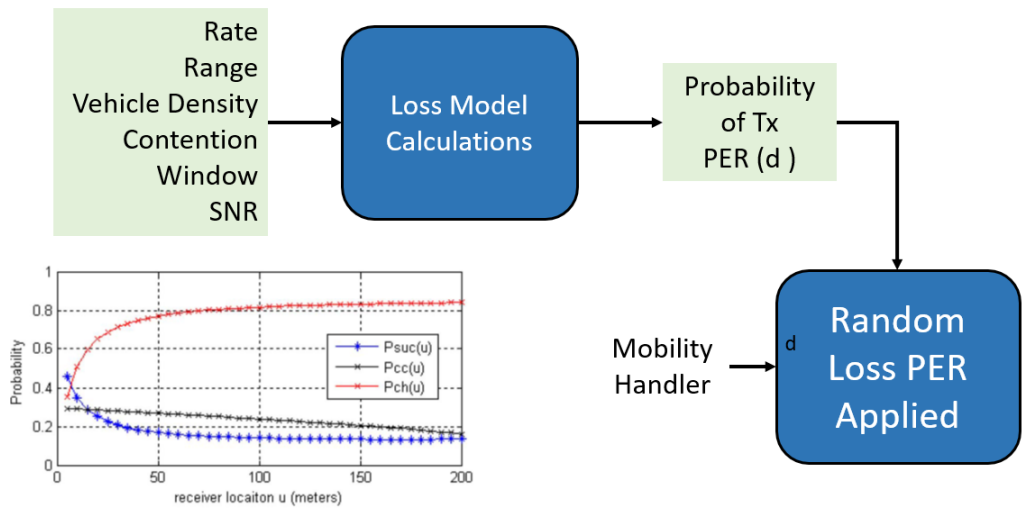

Figure 6-1: Probabilistic Transmission Model

We also plan to add a support for testing Cellular LTE (C-LTE) based V2X applications. This would involve a thorough understanding of several protocols specific to the C-LTE based networks and modifying the underlying architecture to add a cross-platform support. In summary, the RVE has been designed in such a way to incorporate the present and future valuable research around CVS.

# APPENDIX A:
# COMPLEXITY ANALYSIS OF MIN-HEAP

Min-heap is a binary tree-based data structure whose elements are stored in an array. In min-heap, the parent node is either equal to or smaller than its child nodes and the binary tree should be complete. A tree is complete if:

1. Each node except the last two-level nodes must have two child nodes.
2. All the leaf nodes must first occupy the left-most available position in the tree.

A binary tree with n nodes has $\lceil log2(n + 1) \rceil$ levels. An insertion or a deletion requires at most one change at each level. So, either inserting a new node or deleting a node have a time complexity of $O(log2n)$.

Let's take an example:

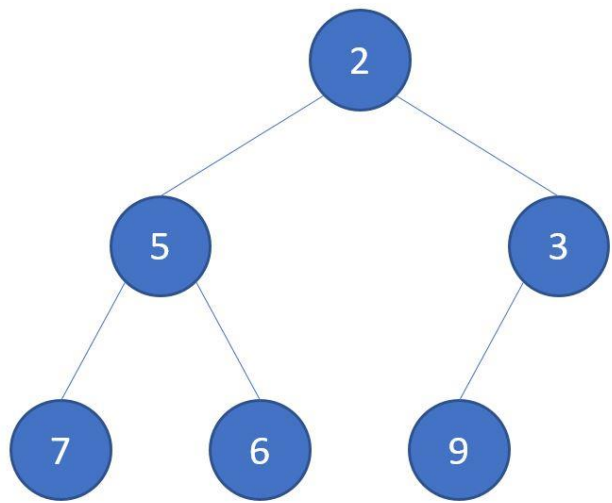

Figure A-1: Example Min-Heap

### A.1 Insertion

To insert new node in the min-heap, add the new node to the end of the tree. Start replacing the parent node with the child node until both the child node's value is either equal to or less than the parent node or the process reaches the root node.

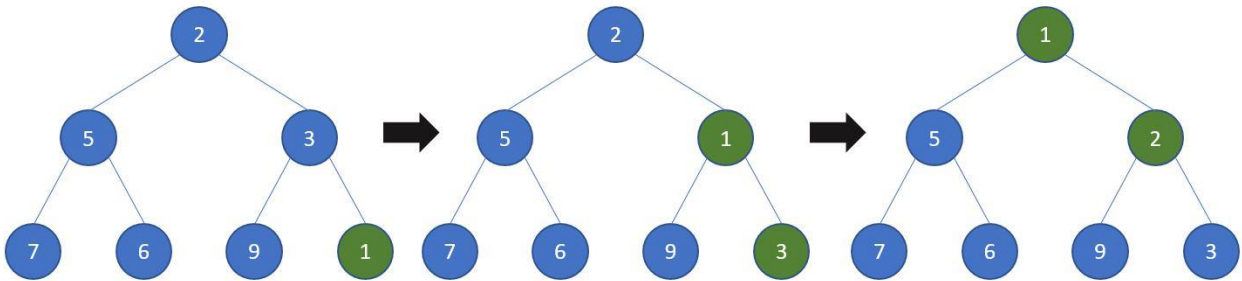

Figure A-2: Min-Heap Insertion

### A.2 Deletion

Fetch the root node value and replace the root node with the last node. Start replacing the parent node with the child node having minimum value. Continue this process until the parent node's value is less than or equal to both the child node's value or the process has reached the leaf node.

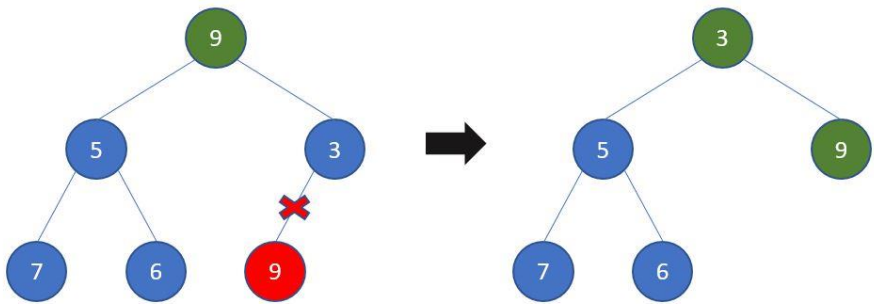

Figure A-3: Min-Heap Deletion